\documentclass[aps,prd,reprint,twocolumn,superscriptaddress,groupedaddress,nofootinbib]{revtex4-1}

\usepackage{changes}
\usepackage{amsmath,amssymb, mathtools}
\usepackage{url}
\usepackage{epsfig}
\usepackage{comment}
\usepackage{graphicx}
\usepackage{slashed}
\usepackage{natbib}
\usepackage{graphicx}
\usepackage[colorlinks=true,linkcolor=blue,citecolor=blue]{hyperref}%
\allowdisplaybreaks

\begin{document}
\title{A new mechanism for matter-antimatter asymmetry and connection with dark matter}

\author{Arnab Dasgupta$^1$, P. S. Bhupal Dev$^2$, Sin Kyu Kang$^1$, Yongchao Zhang$^{2,3}$}
%\email{arnabdasgupta@protonmail.ch}
\affiliation{$^1$School of Liberal Arts, Seoul-Tech, Seoul 139-743, Korea}
%\email{bdev@wustl.edu}
\affiliation{$^2$Department of Physics and McDonnell Center for the Space Sciences, Washington University, St. Louis, MO 63130, USA}
\affiliation{$^3$Department of Physics, Oklahoma State University, Stillwater, OK  74078, USA}
%\email{yongchao.zhang@physics.wustl.edu}

\begin{abstract}

We propose a new mechanism for generating matter-antimatter asymmetry via the interference of tree-level diagrams only, where the imaginary part of the Breit-Wigner propagator for an unstable mediator plays a crucial role. We first derive a general result that a nonzero $CP$-asymmetry can be generated via at least two sets of interfering tree-level diagrams involving either $2\rightarrow2$ or $1\rightarrow n$ (with $n\geq3$) processes. We illustrate this point in a simple  TeV-scale extension of the Standard Model with an inert Higgs doublet and right-handed neutrinos, along with an electroweak-triplet scalar field, where small Majorana neutrino masses are generated via a combination of radiative type-I and tree-level type-II seesaw mechanisms. The imaginary part needed for the required $CP$-asymmetry comes from the trilinear coupling of the inert doublet with the triplet scalar, along with the width of the triplet scalar mediator. The real part of the neutral component of the inert doublet serves as a cold dark matter candidate. The evolutions of the dark matter relic density and the baryon asymmetry are intimately related in this scenario.
\end{abstract}

\maketitle

%{\bf Introduction.--}
\section{Introduction}
The observed asymmetry between the number densities of baryonic matter and antimatter in the universe~\cite{Aghanim:2018eyx} cannot be accounted for in the Standard Model (SM). Therefore, a viable baryogenesis mechanism is an essential ingredient for the success of any beyond SM scenario. The dynamical generation of baryon asymmetry requires three necessary (but not sufficient) Sakharov conditions~\cite{Sakharov:1967dj} to be satisfied: (i) baryon number ($B$) violation, (ii) $C$ and $CP$ violation, and (iii) out-of-equilibrium dynamics. A well-known mechanism that satisfies these conditions involves the $1\to 2$ decays of a heavy particle, such as in grand unified theory (GUT) baryogenesis~\cite{Dimopoulos:1978pw, Weinberg:1979bt} or leptogenesis~\cite{Fukugita:1986hr} (for reviews, see e.g.~Refs.~\cite{Cline:2006ts, Davidson:2008bu}). To obtain a baryon/lepton asymmetry in these
%out-of-equilibrium baryon/lepton number violating
$1\to 2$ decay scenarios, one must consider the interference between tree- and loop-level diagrams. Furthermore, some particles in the loop must be able to go on-shell, and the interaction between the intermediate on-shell particles and the final-state particles should correspond to a net change in baryon/lepton number for the net asymmetry to be nonzero; this is known as the Nanopoulos-Weinberg theorem~\cite{Nanopoulos:1979gx} (see also Refs.~\cite{Adhikari:2001yr, Bhattacharya:2011sy}). Similar interference effects between tree and loop-level diagrams have also been considered for generating the baryon asymmetry from $2\to2$ scattering~\cite{Yoshimura:1978ex, Bento:2001rc, Baldes:2014gca, Baldes:2014rda} or annihilation~\cite{Farrar:2005zd, Gu:2009yx, Cui:2011ab, Kumar:2013uca, Dasgupta:2016odo, Borah:2018uci}  processes.

In this paper, we argue that the interference between tree and loop-level diagrams is not the only way to generate a nonzero asymmetry from out-of-equilibrium heavy particle decays/annihilations.
%To realise baryogenesis or leptogenesis and generate the matter-antimatter asymmetry, one does {\it not} have to resort to the loop-level diagrams.
We propose a new mechanism where it suffices to consider two sets of interfering diagrams at the tree-level only. This can be achieved through  tree-level $2 \to 2$ scattering or $1\to {n}$ (with ${n} \geq 3$) decay processes mediated by unstable particles. Then the $CP$-asymmetry can be generated from the complex couplings and the propagator widths [see Eq.~(\ref{eq:delta}) below], which could even be resonantly enhanced when the center-of-mass energy is close to the propagator mass.
%Without the loops, the leptogenesis scale can in principle be lower than those with loops such as the vanilla leptogenesis. \BD{This needs clarification.}

%In the next section, we discuss the general framework in which this can be realized.

To illustrate our new mechanism, we will consider a simple realistic model at TeV-scale, namely, combining the scotogenic model~\cite{Ma:2006km} (with an inert Higgs doublet and right-handed neutrinos) and the type-II seesaw framework~\cite{Magg:1980ut, Schechter:1980gr, Cheng:1980qt, Mohapatra:1980yp, Lazarides:1980nt} (with an $SU(2)_L$-triplet scalar) for small Majorana neutrino masses. For our parameter choice of the model, the $CP$-asymmetry originates from the complex trilinear coupling of the inert Higgs doublet with the triplet scalar, along with the imaginary part of the triplet scalar mediator width.
%The neutrino masses receive both contributions from the radiative type-I seesaw and tree-level type-II seesaw.
Stabilized by a discrete $Z_2$ symmetry,  the neutral component of the inert doublet scalar plays the role of a TeV-scale weakly-interacting dark matter (DM) candidate. Adopting three benchmark points (BPs), we illustrate that the generation of baryon asymmetry via leptogenesis in this model is intimately correlated with the DM relic density. It is also found that successful leptogenesis sets limits on the triplet vacuum expectation value (VEV) $v_\Delta$ and the trilinear scalar coupling $|\mu_{\eta\Delta}|$ (see Fig.~\ref{fig:scan}) -- a feature that could be directly tested at future high-energy colliders.

The rest of this paper is organized as follows. Our general mechanism of attaining the $CP$ asymmetry without having explicit loop diagrams is explained in Section~\ref{sec:mechanism}. The scotogenic plus type-II seesaw model is introduced in Section~\ref{sec:model}. The lepton asymmetry generation in this model is detailed in Section~\ref{sec:asymmetry}. The collider signatures are touched upon in Section~\ref{sec:collider}.  Our conclusions are given in Section~\ref{sec:conclusion}. The scalar potential and scalar masses are collected in Appendix~\ref{sec:scalars}. The relevant thermal cross sections used in our analysis are given in Appendix~\ref{sec:xsec}. The thermal cross section relevant for the asymmetry in the narrow-width approximation is given in Appendix~\ref{sec:appasym}.

%show  then following that we will show the success of such mechanism by taking a well studied model of Type-II scotogenic model and its result and then we would conclude.

%The main crux of the analysis lies in the technique of generation of asymmetry. We should take a slight detour in explaining the asymmetry.

%{\bf Asymmetry generation without loops.--}
%\section{Mechanism for Asymmetry without loops}
\section{The general mechanism}
\label{sec:mechanism}

We propose that a net lepton or baryon  asymmetry can be generated from the interference effect of two sets of tree-level decay or scattering diagrams with the same initial and final states, as long as the following two conditions are  satisfied:
\begin{enumerate}
\item [(i)]  There is a net nonzero lepton or baryon number between the initial and final states.\footnote{In principle, this condition can be somewhat relaxed if we consider flavor-dependent asymmetries, with zero net lepton or baryon number in the final state, as in flavored leptogenesis (for recent reviews, see e.g. Refs.~\cite{Dev:2017trv, Drewes:2017zyw}). For simplicity, here we will not consider such  flavor-dependent effects.}
\item [(ii)]  At least one set of decay or scattering amplitudes is complex such that the squared amplitudes for particles and antiparticles are different, giving rise to a net $CP$-asymmetry.
\end{enumerate}

\begin{figure}[t!]
    \centering
\includegraphics[width=0.4\textwidth]{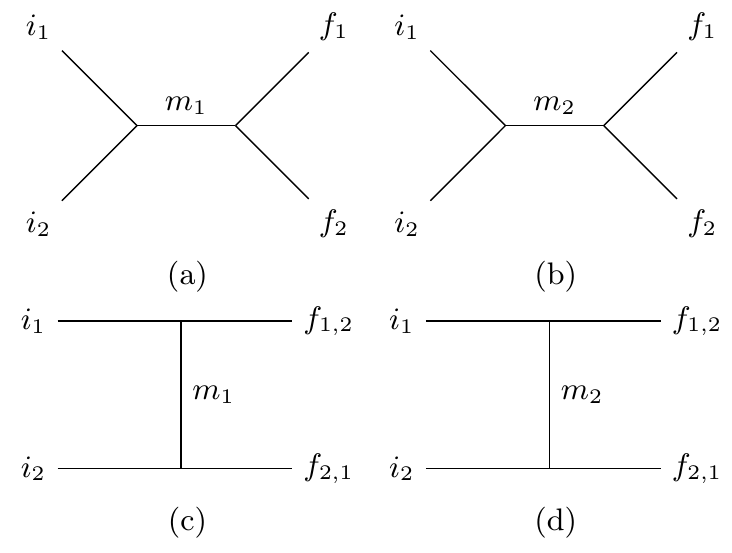}
\caption{Generic topologies for tree-level $2\rightarrow2$ subprocesses that can give rise to a nonzero lepton or baryon asymmetry. Here $i_{1,2}$ and $f_{1,2}$ are respectively the initial and final states, and $m_{1,2}$ are the masses of two different mediators. }
\label{fig:woloops22}
\end{figure}

\begin{figure}[t!]
    \centering
\includegraphics[scale=1,trim={6.4cm 22.6cm 6.6cm 4cm},clip,width=0.4\textwidth]{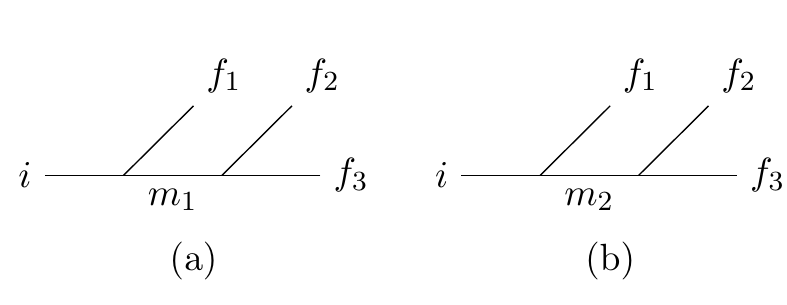}
\caption{Generic topologies for tree-level $1\rightarrow3$ subprocesses that can give rise to a nonzero lepton or baryon asymmetry. Here $i$ and $f_{1,2,3}$ are respectively the initial and final states, and $m_{1,2}$ are the masses of two different mediators. }
\label{fig:woloops13}
\end{figure}

The simplest way to achieve this is through $2\to 2$ scatterings (see Fig.~\ref{fig:woloops22}) or $1\to 3$ decays (see Fig.~\ref{fig:woloops13}) involving two {\it different} intermediate state particles,
%For the interfering of tree and loop level diagram such as in vanilla leptogenesis,
%One elegant example is leptogenesis where lepton asymmetry is generated from the decays of heavy neutrinos $N$, and the two sets of diagrams are respectively the tree and loop level two-body decays $N \to LH,\, \bar{L}H^\dagger$, with $L$ and $H$ respectively the SM lepton and Higgs doublet and $\bar{L}$ and $H^\dagger$ the corresponding antiparticles. In the leptogenesis paradigm,
%the imaginary parts are from the loop-level effects. In fact, the necessary imaginary components for lepton or baryon asymmetry generation can also arise at the tree-level, which is just the crux we are trying to address in this paper. The simplest scenario
%for generation of matter asymmetry
%at the tree-level are through the $2\to2$ scattering processes or $1\to {n}$ decays with ${n} \geq3$,
with the outgoing particles (or decay products) carrying a net nonzero baryon or lepton number. Without loss of generality, we focus here on the simplest $2\to2$ scattering case with the initial states  $i_1, \: i_2$ and with only two subprocesses for the final states $f_1, \: f_2$ (here $i_{1,2}$ and $f_{1,2}$ generically stand for bosons and/or fermions), mediated by intermediate-state particles of mass $m_1$ and $m_2$, respectively.
%see Fig.~\ref{fig:woloops22}. % and the description below.
The total amplitude for the process $i_1 i_2 \to f_1 f_2$ can be written as
%Now, let us assume the process violating $B$or $L$ as ($f\rightarrow p$) where '$p$' carries the baryon/lepton number.
\begin{align}
  \mathcal{M} & \ = \ (\mathcal{C}_1 \mathcal{M}_1 + \mathcal{C}_2 \mathcal{M}_2 ) \mathcal{W}\,,
\end{align}
where $\mathcal{C}_i$ contain only the couplings, $\mathcal{W}$ contains the wave functions for the incoming and outgoing particles and $\mathcal{M}_i$ stand for the rest of the sub-amplitudes. The corresponding amplitude for the conjugate process $\bar{i}_1 \bar{i}_2 \to \bar{f}_1 \bar{f}_2$ is\footnote{Note that the $CPT$ theorem only guarantees the equivalence of rates for $i_1i_2\to f_1f_2$ and $\bar{f_1}\bar{f}_2\to \bar{i}_1\bar{i_2}$.}
%the {\it anti-particles} $\bar{f}_1\bar{f}_2$ (with the same initial state $i_1i_2$) is
%Similarly, if we write the amplitude for the anti-particle it is given as
\begin{align}
\label{eq:amp}
    \overline{\mathcal{M}}  \ = \ (\mathcal{C}^*_1 \mathcal{M}_1 + \mathcal{C}^*_2 \mathcal{M}_2) \mathcal{W}^* \,.
\end{align}
Comparing the modular squares of the amplitudes, we  obtain the $CP$-asymmetry factor
%And hence the asymmetry is given as
\begin{align}
\delta & \ \equiv \ |\mathcal{M}|^2 - |\overline{\mathcal{M}}|^2 \nonumber \\
%&= (\mathcal{C}^*_1\mathcal{M}^*_1 + \mathcal{C}^*_2\mathcal{M}^*_2)(\mathcal{C}_1\mathcal{M}_1 + \mathcal{C}_2\mathcal{M}_2) \nonumber \\
%&- (\mathcal{C}_1\mathcal{M}^*_1 + \mathcal{C}_2\mathcal{M}^*_2)(\mathcal{C}^*_1\mathcal{M}_1 + \mathcal{C}^*_2\mathcal{M}_2) \nonumber \\
& \ = \ -4 \:
{\rm Im} [\mathcal{C}_1 \mathcal{C}_2^\ast]\:
{\rm Im} [\mathcal{M}_1 \mathcal{M}^\ast_2] |\mathcal{W}|^2,
\label{eq:delta}
\end{align}
where ${\rm Im} [\mathcal{C}_1 \mathcal{C}^\ast_2]$ is the imaginary part coming from the couplings, which is required to be nonzero for $CP$ violation, and ${\rm Im} [\mathcal{M}_1 \mathcal{M}_2^\ast]$ incorporates the imaginary part from the sub-amplitudes ${\cal M}_{1,\,2}$, which is reminiscent of the imaginary part coming from the interference of tree and loop-level diagrams in the $1\to 2$ decay scenario.
%Note that for  $1\to 2$ decays, the tree-level decay rates for $f_1f_2$ and $\bar{f}_1\bar{f}_2$ final states are both proportional to the modular square of the same couplings, which implies $\delta=0$ \blue{at tree-level}.
%Therefore, complex couplings are not sufficient (although necessary for $CP$ violation) for realizing $\delta\neq 0$.
Eq.~\eqref{eq:delta} is a general result  applicable to $2\to 2$ scatterings, as well as $1\to {n}$ decays (for ${n}\geq 3$).
%Therefore, one must consider the interference between tree and loop diagrams to generate a nonzero $CP$ asymmetry in the $1\to 2$ decays.

In the tree-level $2\to2$ processes shown in Fig.~\ref{fig:woloops22},
we have only one source for the complex sub-amplitudes, namely, the finite widths of the mediators. One may argue that the finite width is also a loop-induced effect for unstable mediators, since it is related to the imaginary part of self-energy~\cite{Sirlin:1991fd, Grassi:2000dz}. However, the crux of our new mechanism is that we {\it only} require a nonzero width, whereas the $1\to 2$ decay case needs both nonzero width {\it and} interference between tree and loop (self-energy and/or vertex correction)  diagrams.
%s and exists in all the $s$-, $t$- and $u$-channel diagrams.
In general, the sub-amplitudes ${\cal M}_{1,\,2}$ for the processes in Fig.~\ref{fig:woloops22} can be written as
%${\cal M}_{j} = A_{j}/(x_j -m_{j}^2 + i m_j \Gamma_j)$,
%just the resonance effect in the $s$-channel where one have to take into consideration the widths of mediators.
%If both ${\cal M}_{1,\,2}$ are in the $s$-channel, they can be written in the form of
\begin{eqnarray}
{\cal M}_{j} \ = \ \frac{A_{j}}{x_j -m_{j}^2 + i m_j \Gamma_j} \,,
\end{eqnarray}
with $j=1,\,2$, $x_j = s,\, t,\, u$ the Mandelstam variables, $m_{j}$ and $\Gamma_j$ respectively the mediator masses and widths, and $A_j$ some arbitrary real parameters. The imaginary component of the product of sub-amplitudes appearing in Eq.~\eqref{eq:delta} can then be written as
%\begingroup\makeatletter\def\f@size{9}\check@mathfonts
\begin{align}
\label{eqn:ImM1M2}
{\rm Im} [{\cal M}_1 {\cal M}_2^\ast] =
\frac{A_1 A_2 \left[ (x_1-m_1^2)m_2\Gamma_2 - (x_2-m_2^2)m_1\Gamma_1 \right]}{\left[(x_1-m^2_1)^2 + m^2_1\Gamma^2_1\right]\left[(x_2-m^2_2)^2 + m^2_2\Gamma^2_2\right]}, %\nonumber \\ &&
\end{align}
%\endgroup
which is non-vanishing as long as $(x_1-m_1^2)m_2\Gamma_2 \neq (x_2-m_2^2)m_1\Gamma_1$ in the numerator. With the imaginary part of the couplings ${\rm Im}[{\cal C}_1 {\cal C}_2^\ast] \neq 0$, we can then produce a nonzero asymmetry [cf.~Eq.~\eqref{eq:delta}]. This general argument holds, irrespective of the specific  subprocesses or the model details. %The expression in Eq.~(\ref{eqn:ImM1M2}) applies also to the $1\to3$ decay case.

%For the tree-level topologies shown in Fig.~\ref{fig:woloops22},
For the tree-level $2\to2$ case in Fig.~\ref{fig:woloops22}, we can have three distinct possibilities for the two subprocesses to realize ${\rm Im} [{\cal M}_1 {\cal M}_2^\ast]\neq 0$ in Eq.~\eqref{eqn:ImM1M2}:
%Suppose the two $s$-channel mediators are different, i.e. $m_1 \neq m_2$ and $\Gamma_1 \neq \Gamma_2$, and furthermore
%As long as  in Eq.~(\ref{eqn:ImM1M2}), then with the imaginary couplings ${\rm Im}[{\cal C}_1 {\cal C}_2^\ast] \neq 0$, we can produce nonzero asymmetry $\delta \neq 0$. This is the most general argument and does not depend on the specific channels of the two subprocesses or the model details. The expression in Eq.~(\ref{eqn:ImM1M2}) applies also to the $1\to3$ decay case.
\begin{enumerate}
\item [(i)] If both subprocesses are in the $s$-channel [cf.~Fig.~\ref{fig:woloops22} (a)+(b)],
one just needs to replace $x_{1,\,2}$ by $s$ in Eq.~(\ref{eqn:ImM1M2}). In this case,  the $CP$-asymmetry factor $\delta$ in Eq.~\eqref{eq:delta} can be largely enhanced in the vicinity of resonance(s), with $s-m_i^2 \simeq m_i \Gamma_i$ (with $i = 1,\,2$), similar to the enhancement effect in resonant leptogenesis~\cite{Pilaftsis:2003gt, Dev:2017wwc}.

\item [(ii)] If one of the sub-amplitudes is in the $s$-channel and the other one in the $t$- or $u$-channel [cf.~Fig.~\ref{fig:woloops22} (a)+(d) or (b)+(c)],
one can safely neglect the imaginary part for the $t$- or $u$-channel propagator. For concreteness, we take ${\cal M}_1$ as the $s$-channel and ${\cal M}_2$ as the $x$-channel ($x=t$ or $u$) amplitude. In this case, Eq.~\eqref{eqn:ImM1M2} can be simplified to
\begin{eqnarray}
{\rm Im} [{\cal M}_1 {\cal M}_2^\ast] \ \simeq \ -
\frac{A_1 A_2 m_1\Gamma_1}{\left[(s-m^2_1)^2 + m^2_1\Gamma^2_1\right](x-m^2_2)} \,,
\label{eq:6}
\end{eqnarray}
which is proportional to the $s$-channel mediator width $\Gamma_1$. Here also the $CP$-asymmetry could be largely enhanced at the $s$-channel resonance, i.e. $s-m_1^2 \simeq m_1 \Gamma_1$.

\item [(iii)] If both subprocesses are in the $t$- or $u$-channel [cf.~Fig.~\ref{fig:woloops22} (c)+(d)],
then the width terms in the denominator of Eq.~(\ref{eqn:ImM1M2}) can be neglected, i.e.
%\begingroup\makeatletter\def\f@size{9}\check@mathfonts
\begin{align}
\label{eqn:ImM1M2simp}
{\rm Im} [{\cal M}_1 {\cal M}_2^\ast] \simeq
\frac{A_1 A_2 \left[ (x_1-m_1^2)m_2\Gamma_2 - (x_2-m_2^2)m_1\Gamma_1 \right]}{(x_1-m^2_1)^2 (x_2-m^2_2)^2}.
\end{align}
%\endgroup
In this case, the $CP$-asymmetry is suppressed by the ratio $m_i\Gamma_i/(x_j-m_j^2)$ with $i,\, j = 1,\,2$.
\end{enumerate}
In the next section, we will consider an explicit example that realizes the possibility (ii) discussed above.

\section{An Example}
\label{sec:model}

To illustrate our new mechanism in a minimal realistic extension of SM, we consider an amalgamation of the scotogenic model~\cite{Ma:2006km} and type-II seesaw~\cite{Magg:1980ut, Schechter:1980gr, Cheng:1980qt, Mohapatra:1980yp, Lazarides:1980nt} mechanisms at TeV-scale. For the purpose of scotogenic mechanism, an inert $SU(2)_L$-doublet scalar $\eta\equiv (\eta^+, \eta^0)$ and three right-handed neutrinos (RHNs) $N_i$ (with $i=1,2,3$) are introduced. To implement type-II seesaw, an $SU(2)_L$-triplet scalar $\Delta\equiv (\Delta^{++}, \Delta^+,\Delta^0)$ is added.
%Written in the component form, these beyond SM scalars read,  with the quantum numbers under the SM $SU(2)_L \times U(1)_Y$ group put in the parenthesis:
%\begingroup\makeatletter\def\f@size{9}\check@mathfonts
%\begin{eqnarray}
%\eta =\begin{pmatrix}\eta^+ \\ \eta^0\end{pmatrix} \in ({\bf 2},\,1/2) \,, \;\;
%\Delta = \begin{pmatrix}\Delta^{++} & \Delta^+ \\
%\Delta^+ & -\Delta^0\end{pmatrix} \in ({\bf 3},\, 1) \,.
%\end{eqnarray}
%\endgroup
The inert doublet $\eta$ and the three RHNs $N_i$ are odd under a discrete $Z_2$ symmetry, while all other particles are even. In this model, we assume the RHNs, as well as the triplet scalar components,  are heavier than the $\eta$ scalars, so any asymmetry generated by the conventional decays of $N$ and/or $\Delta$ is not relevant at the temperature scale of interest. An added advantage of our mechanism is that the lightest neutral component $\eta^0$ plays the role of DM~\cite{Ma:2006km}, with its relic density intimately connected to the lepton asymmetry. A nonminimal coupling of the inert doublet to gravity can also successfully accommodate  inflation~\cite{Choubey:2017hsq, Borah:2018rca}.

The relevant Yukawa couplings are given by the Lagrangian
\begin{eqnarray}
\label{eqn:Yukawa}
-\mathcal{L}_Y \ = \ Y^N_{i\alpha} \widetilde{\eta}^\dagger L_\alpha N_i + Y^\Delta_{\alpha \beta} \overline{L_\alpha^C} \Delta L_\beta + {\rm H.c.}\,,
\end{eqnarray}
with $L\equiv (\nu, \ell)$ being the SM lepton doublet, $C$ the charge conjugation operator, $\widetilde\eta = i\sigma_2 \eta^\ast$ ($\sigma_2$ being the second Pauli matrix), $\alpha,\, \beta = e,\, \mu,\, \tau$ the lepton flavor indices, and $i = 1,\, 2,\, 3$ the RHN mass indices. For simplicity, we assume there is no mixing nor $CP$ violation in the RHN sector.
The mass parameter $\mu_{\eta \Delta}$
in the scalar potential
\begin{eqnarray}
\label{eqn:trilinear}
V \ \supset \ \mu_{\eta \Delta} \eta^\dagger \Delta^\dagger \widetilde{\eta} + {\rm H.c.}
\end{eqnarray}
is chosen to be complex, which is crucial for the $CP$-asymmetry [cf.~Eq.~(\ref{eq:delta})].
%All other parameters in Eq.~(\ref{eqn:potential}) are assumed to be real.
The full scalar potential and the resultant physical scalar masses are collected in Appendix~\ref{sec:scalars}.
%, with the real component $\eta_R$ of $\eta^0$ being the DM candidate.
%for the neutral and charged scalars can be obtained from the minimization of the potential (\ref{eqn:potential}), which is detailed in the Appendix.
%Note that here the doublet $\eta$ is odd under the $Z_2$ symmetry and does not mix the SM Higgs and the triplet, which is necessary for the neutral real component $\eta_{\rm R}$ to be a stable DM candidate.

In this setup, the neutrino mass is generated from both loop-level scotogenic and tree-level type-II seesaw mechanisms, which are induced respectively by the Yukawa couplings $Y^N$ and $Y^\Delta$ given in Eq.~(\ref{eqn:Yukawa}):
%\blue{\begin{eqnarray}
%\label{eqn:FI}
%F_{\rm I} m_\nu & \ \equiv \ & (Y^{N})^{\sf T} \Lambda Y^N_{} \,, \\
%\label{eqn:FII}
%F_{\rm II} m_\nu & \ \equiv \ & Y^\Delta_{} v_\Delta \,,
%\end{eqnarray}}
\begin{eqnarray}
\label{eqn:mnu}
m_{\nu} \ = \
(Y^{N})^{\sf T} \Lambda Y^N_{} + Y^\Delta_{} v_\Delta \, ,
\end{eqnarray}
where $\Lambda$ is an effective loop-suppressed RHN mass scale, given by~\cite{Ma:2006km, Merle:2015ica}
\begin{eqnarray}
\Lambda_{ii} & \ = \ & \frac{m_{N_i}}{16\pi^2} \left[\frac{m^2_{\eta_{R}}}{m^2_{N_i}-m^2_{\eta_{R}}}\ln \left(\frac{m^2_{N_i}}{m^2_{\eta_{R}}}\right) \right. \nonumber \\
&& \quad\qquad \left. - \frac{m^2_{\eta_{\rm I}}}{m^2_{N_i}-m^2_{\eta_{I}}}\ln \left(\frac{m^2_{N_i}}{m^2_{\eta_{I}}}\right)\right] \, .
\end{eqnarray}
We have assumed that the RHNs do not mix with each other, therefore $\Lambda$ is a diagonal matrix. The Yukawa couplings in Eq.~(\ref{eqn:mnu}) are related to the neutrino oscillation data, $\Lambda$ and the triplet VEV $\langle \Delta^0 \rangle=v_\Delta$ as follows:
\begin{align}
\label{eqn:YN}
    Y^N_{i\alpha} & \ = \ F_{\rm I}^{1/2} \left(\Lambda^{-1/2} {\cal O} \widehat{m}^{1/2}_\nu U^\dagger_{\rm PMNS}\right)_{i\alpha} \,, \\
    \label{eqn:YDelta}
    Y^\Delta_{\alpha \beta} & \ = \ F_{\rm II} v_{\Delta}^{-1} (U^*_{\rm PMNS} \widehat{m}_\nu U^\dagger_{\rm PMNS})_{\alpha \beta} \,,
\end{align}
where $\widehat{m}_\nu = \{ m_{\nu_1},\, m_{\nu_2},\, m_{\nu_3} \}$ the diagonal neutrino mass eigenvalues, and $U_{\rm PMNS}$ the PMNS lepton mixing matrix. In Eq.~(\ref{eqn:YN}) we have used the Casas-Ibarra parametrization~\cite{Casas:2001sr} for the coupling $Y^N$, with ${\cal O}$ an arbitrary orthogonal matrix. $F_{\rm I}$ and $F_{\rm II}$ are the fractions of contributions to neutrino mass matrix from the radiative scotogenic and tree-level type-II seesaw mechanisms respectively, with $F_{\rm I} + F_{\rm II} = 1$.

%\section{lepton asymmetry generation}

%{\bf Boltzmann equations.--}
\section{Generation of lepton asymmetry and Dark Matter relic density}
\label{sec:asymmetry}

\begin{figure}[t!]
\centering
\includegraphics[width=0.4\textwidth]{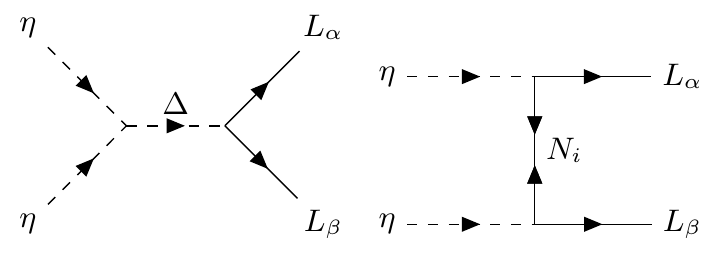}
\caption{Feynman diagrams for the $2\to 2$ scattering processes $\eta\eta\to L_\alpha L_\beta$ in our example model.}
\label{fig:asymfeyn}
\end{figure}

As stated above, the matter asymmetry is generated from the interference effect between two tree-level diagrams, which are shown in Fig.~\ref{fig:asymfeyn} for our scotogenic type-II seesaw model with $m_\Delta\gtrsim 2m_\eta$.\footnote{In the opposite regime where $m_\Delta<2m_\eta$, an asymmetry can be generated via the interference between tree-level $\Delta\to LL$ decay mediated by the Yukawa coupling $Y^\Delta$ and the vertex correction to this decay induced by two $\eta$'s and a $N$ mediated by the couplings $\mu_{\eta \Delta}$ and $Y^N$ respectively.} In particular, we analyze the $2\to 2$  $\Delta L=2$ scattering processes
\begin{eqnarray}
\label{eqn:process}
%(\eta^-,\eta^0) (\eta^-,\eta^0) \to (\ell_\alpha^-,\nu_\alpha) (\ell_\beta^-,\nu_\beta) \,.
%\eta^\pm \eta^\pm \to \ell_\alpha^\pm \ell_\beta^\pm \,.
\eta \eta \ \to \ L_\alpha L_\beta \,,
\end{eqnarray}
which include $\eta^\pm \eta^\pm \to \ell_\alpha^\pm \ell_\beta^\pm$, $\eta^0 \eta^\pm \to \ell_\alpha^\pm \nu_\beta$ and $\eta^0 \eta^0 \to \nu_\alpha \nu_\beta$. These processes can be mediated by an $s$-channel triplet scalar  $\Delta$, and also by RHNs $N_i$ in the $t$- and $u$-channels, as shown in Fig.~\ref{fig:asymfeyn}.
%In these processes we can produce either two units or minus two units of lepton numbers. Then it is straightforward to get
The effective $CP$-asymmetry factor [cf.~Eq.~\eqref{eq:6}] is given by
\begin{align}
\label{eqn:delta2}
\delta & \ = \ 4
\sum_{i} \: {\rm Im}
\left[\mu_{\eta \Delta} \big\{Y^N Y^{\Delta^*}(Y^{N})^{\sf T} \big\}_{ii} \right] \nonumber \\
&\times \frac{s m_{N_i}  m_{\Delta} \Gamma_{\Delta^{}}}{(s-m^2_{\Delta})^2 + m^2_{\Delta^{}}\Gamma^2_{\Delta^{}}}
    \left[\frac{1}{t-m^2_{N_i}} + \frac{1}{u-m^2_{N_i}}\right] \, ,
\end{align}
where $\Gamma_{\Delta^{}}$ is the triplet scalar width. With the width in the numerator of Eq.~(\ref{eqn:delta2}), the asymmetry in this simple model can be viewed as the
interference effect of RHN-mediated tree-level diagram in the right panel of Fig.~\ref{fig:asymfeyn} and the one-loop correction to the triplet propagator in the $s$-channel, where the $s$-channel process  corresponds to the following subprocesses
\begin{eqnarray}
\eta\eta \to \Delta^{(\ast)}\,, \quad
\Delta^{(\ast)} \to L L
\end{eqnarray}
which incorporate the conventional one-loop decay of (on-shell) triplet scalars into a lepton pair.

From an effective field theory (EFT) perspective, if we integrate out the heavy mediator masses $m_{\Delta}$ and $m_{N_i}$, the width effect has to be consistently incorporated into the effective coupling. This can be understood by inserting a self-energy diagram in the $\Delta$-propagator in Fig.~\ref{fig:asymfeyn} and then integrating out the resulting two $\Delta$ propagators to obtain a loop-level effective coupling that includes the width of $\Delta$, giving rise to a non-vanishing asymmetry.
%This can be understood as a loop-effect, where  we get the following expression:
%\begin{align}
 %   \delta &\simeq \
%\sum_{i} 4\: {\rm Im}
%\left[\mu_{\eta \Delta} \big\{Y^N Y^{\Delta^*}(Y^{N})^{\sf T} \big\}_{ii} \right] \nonumber \\
%&\times \frac{s\Gamma_{\Delta^{}}}{m^3_{\Delta}m_{N_i}}.
%\end{align}

In Eq.~\eqref{eqn:delta2}, the imaginary part of the product of the Yukawa couplings $Y^N$, $Y^\Delta$ [cf.~Eq.~\eqref{eqn:Yukawa}] and the trilinear coupling $\mu_{\eta \Delta}$ [cf.~Eq.~\eqref{eqn:trilinear}] can be parameterized using Eqs.~\eqref{eqn:YN} and \eqref{eqn:YDelta} as follows:
%\begin{widetext}
%\begin{align}
%    \sum_{\alpha \beta} \Im[\mu_{\eta \Delta}Y^N_{i\alpha}Y^{\Delta^*}_{\alpha \beta}Y^N_{i\beta}] &= \Im[\mu_{\eta \Delta} Tr[Y^NY^{\Delta^*}Y^{N^T}]] \nonumber \\
%    &= \Im[\mu_{\eta \Delta}Tr[\sqrt{\Lambda}^{-1}R\sqrt{m^{diag}_\nu}U^\dagger_{PMNS}U_{PMNS}m^{diag}_\nu U_{PMNS}^TU_{PMNS}^*\sqrt{m^{diag}_\nu}R^T\sqrt{\Lambda}^{-1}]] \nonumber \\
%    &= \frac{\Im[\mu_{\eta \Delta}]}{\langle \Delta^0 \rangle}Tr[(m^{diag})^2\Lambda^{-1}].
%\end{align}
%\end{widetext}
\begin{align}
\label{eqn:imaginary}
    & {\rm Im}
\left[\mu_{\eta \Delta} \big\{ Y^N Y^{\Delta^*}(Y^{N})^{\sf T} \big\}_{ii} \right] \nonumber \\
    & = F_{\rm I} F_{\rm II} v_{\Delta}^{-1}
    {\rm Im} \left[ \mu_{\eta \Delta} \big\{
    \Lambda^{-1/2} {\cal O} \, \widehat{m}_\nu^2 {\cal O}^{\sf T} \Lambda^{-1/2} \big\}_{ii} \right] \,.
\end{align}
In general, the orthogonal ${\cal O}$-matrix might be complex, thus potentially contributing to the imaginary part in Eq.~(\ref{eqn:imaginary}). % and the asymmetry factor in Eq.~(\ref{eqn:delta2}).
%and ${\cal O} \widehat{m}_\nu^2 {\cal O}^{\sf T} \neq \widehat{m}_\nu^2$ if ${\cal O}$ is not an identity matrix.

It is interesting that part of the same $2\to 2$ processes in Eq.~\eqref{eqn:process} containing $\eta^0$ contributes also to the (co)annihilation of DM particles. In this sense, the time evolutions of DM relic density and the lepton asymmetry are correlated, as we will see below. The freeze-out mechanism for DM is identical to the standard inert doublet case~\cite{Barbieri:2006dq,LopezHonorez:2006gr}, where DM annihilates into the SM particles.
%we can have $\eta \eta \rightarrow {\rm SM~SM}$ with ``SM'' standing for all SM fermions (quarks and leptons), gauge bosons $W,Z$ and Higgs boson $h$.

\subsection{Boltzmann equations}
\label{sec:equations}

The cogenesis of DM relic density and leptonic asymmetry is governed by the coupled Boltzmann equations
\begin{align}
\label{eqn:Boltzmann}
\frac{{\rm d}Y_\eta}{{\rm d}z}  \ = \ &
\frac{-s}{H(z)z}\left[(Y^2_\eta - (Y^{\rm eq}_\eta)^2)\langle \sigma v \rangle(\eta \eta \rightarrow {\rm SM} \, {\rm SM}) \right] ,  \\
\frac{{\rm d}Y_{\Delta L}}{{\rm d}z} \ = \ &
\frac{s}{H(z)z}\left[(Y^2_\eta - (Y^{\rm eq}_\eta)^2)\langle \sigma v \rangle_{\delta} (\eta \eta \rightarrow L L) \right. \nonumber \\
&-  2Y_{\Delta L}Y^{\rm eq}_\ell r^2_{\eta} \langle \sigma v \rangle_{\rm tot}(\eta \eta \rightarrow L L) \nonumber \\
&- \left. 2Y_{\Delta L}Y^{\rm eq}_\eta \langle \sigma v \rangle( \eta \bar{L} \rightarrow \eta L)\right] \,, \label{eqn:Boltzmann2}
\end{align}
where $z = m_\eta/T$, $Y_i^{(\rm eq)}\equiv n_i^{(\rm eq)}/s$ are the normalized number densities (in equilibrium) for the particles $i$ ($s$ being
the entropy density), $Y_{\Delta L} = Y_L - Y_{\bar{L}}$, $r_\eta = Y^{\rm eq}_\eta/Y^{\rm eq}_\ell$,  and
\begin{eqnarray}
H(z) \  = \  \sqrt{\frac{8\pi^3 g_\ast}{90}} \frac{m_{\eta}^2}{z^2M_{\rm Pl}}
\end{eqnarray}
%$H(z) = \sqrt{8\pi^3 g_\ast/90}\: m_{\eta}^2/(z^2M_{\rm Pl})$
with $M_{\rm Pl}$ the Planck scale and $g_\ast$ the number of relativistic degrees of freedom at temperature $T$. Here the $\langle\sigma v\rangle$'s are the thermally-averaged annihilation/scattering rates: $\langle \sigma v \rangle (\eta \eta \to {\rm SM~SM})$ is the DM
annihilation rate, and $\langle \sigma v \rangle_{\rm tot,\,\delta} (\eta \eta \to LL)$ are  respectively given by
%is the thermally-averaged scattering cross section for $\eta\eta \to LL$ that includes also the $\Delta L=0$ processes such as $\eta^+ \eta^- \to \ell^+ \ell^-$,
\begin{eqnarray}
\langle \sigma v \rangle_{\rm tot} (\eta \eta \to LL) & \ \equiv \ &
\langle \sigma v \rangle_{} (\eta \eta \to LL) \nonumber \\
&& + \langle \sigma v \rangle_{} (\eta^\ast \eta^\ast \to \bar{L}\bar{L}) \,,  \\
\langle \sigma v \rangle_{\delta} (\eta \eta \to LL) & \ \equiv \ &
\langle \sigma v \rangle_{} (\eta \eta \to LL) \nonumber \\
&& - \langle \sigma v \rangle_{} (\eta^\ast \eta^\ast \to \bar{L}\bar{L}) \,.
\end{eqnarray}
%\blue{includes the thermally-averaged scattering cross section for both $\eta\eta \to LL$ and $\eta^\ast \eta^\ast \to \bar{L} \bar{L}$},
%whereas $\langle \sigma v \rangle_{\delta} (\eta \eta \to LL)$ includes only the $\Delta L=2$ processes listed in Eq.~\eqref{eqn:process} and the amplitude for which is given in Eq.~\eqref{eqn:delta2}.
The expressions for all the thermal cross sections appearing in Eqs.~(\ref{eqn:Boltzmann}) and (\ref{eqn:Boltzmann2}) are collected in Appendices~\ref{sec:xsec} and \ref{sec:appasym}.

Evaluating the Boltzmann equations above, one can obtain the lepton asymmetry $Y_{\Delta L}(z)$, which is then converted to baryon asymmetry $Y_{\Delta B}=-(28/51)Y_{\Delta L}$~\cite{Harvey:1990qw} via the standard electroweak sphaleron processes~\cite{Kuzmin:1985mm} at the sphaleron transition temperature $T_{\rm sph} = (131.7\pm 2.3)$ GeV~\cite{DOnofrio:2014rug}. In an analogous way, one can also calculate the evolution of the DM density $Y_\eta$ from Eq.~\eqref{eqn:Boltzmann} and get the final relic abundance $\Omega_{\rm DM}h^2=2.755\times 10^8 Y_\eta (m_\eta/{\rm GeV})$ at DM freeze-out temperature $T_f\simeq m_\eta/20$.

We note here that our mechanism for generating the lepton asymmetry and DM relic density simultaneously is similar to the WIMPy baryogenesis mechanism~\cite{Cui:2011ab}. A crucial criterion for achieving successful asymmetry in both cases is that the washout of the asymmetry processes must freeze-out before the freeze-out of the DM annihilation processes, i.e~$\langle \sigma v \rangle_{\rm tot} (\eta \eta \to LL)<\langle \sigma v \rangle(\eta \eta \rightarrow {\rm SM} \, {\rm SM})$. In Ref.~\cite{Cui:2011ab}, both washout and DM freeze-out are governed by the same final states; therefore, one of the final states is required to be massive to satisfy the above freeze-out condition. In our case, however, the dominant process for DM freeze-out is $\eta\eta\to W^+W^-$ via $SU(2)_L$ gauge interaction, whereas the dominant washout process is $\eta\eta\to LL$ via the Yukawa couplings; therefore, we can satisfy the freeze-out condition for suitable choice of the Yukawa couplings without requiring any of the final states to be massive.

%\subsection{Benchmark points}
%\section{Benchmark points}

%{\bf Numerical results.--}
\subsection{Numerical results}
\label{sec:result}

\begin{table}[!t]
    \centering
    \caption{Three BPs for the numerical analysis. All the quartic couplings in Eq.~(\ref{eqn:potential}) not listed in this table are set to be zero. Here $\Delta m_{\eta^0} = m_{\eta_R}-m_{\eta_I}$ is the mass splitting between the two scalars $\eta_{R}$ and $\eta_{I}$.}
    \label{tab:BP2}
    \begin{tabular}{cccc} \hline\hline
    %\multicolumn{4}{|c|}{Parameters/Masses} \\\hline
     & BP1  & BP2 & BP3\\ \hline
    $v_\Delta$ & $1$  keV & $1$  keV & $1$  keV \\ \hline
    $\mu_\eta$  & 600 GeV & 1 TeV & 1.5 TeV\\ \hline
    $\mu_{H\Delta}$ & $33.6$ keV & $93.5$ keV & $210$ keV\\ \hline
    $\mu_{\eta \Delta}$ & 15$i$ GeV & 7.1$i$ GeV & 6$i$ GeV\\ \hline
    $m_{N_1}$ & 6 TeV  &  10 TeV & 15 TeV\\ \hline
    $m_{N_2}$ & 6.6 TeV  & 11 TeV & 16.5 TeV\\ \hline
    $m_{N_3}$ & 7.2 TeV  & 12 TeV & 18 TeV\\ \hline
    $m_{\eta^0}$ & 600 GeV & 1 TeV & 1.5 TeV\\ \hline
    $\Delta m_{\eta^0}$ & $506$ keV & $300$ keV & $200$ keV\\ \hline
    $m_{\eta^\pm}$ & 606 GeV & 1 TeV & 1.5 TeV\\ \hline
    $m_{\Delta^0_{}}$ & 1.2 TeV & 2 TeV & 3 TeV\\ \hline
    $m_{\Delta^\pm}$ & 1.2 TeV & 2 TeV & 3 TeV\\ \hline
    $m_{\Delta^{\pm\pm}}$  & 1.2 TeV & 2 TeV & 3 TeV\\ \hline
    $\lambda_H$ & 0.253 & 0.253 & 0.253 \\ \hline
    $\lambda_{H\eta}$ & 0.19  & 0.56 & 0.91\\ \hline
    $\lambda^\prime_{H\eta}$ & $-0.19$  & $-0.56$ & $-0.91$ \\ \hline
    $\lambda^{\prime \prime}_{H\eta}$ & $1\times10^{-5}$ & $1\times10^{-5}$ & $1\times10^{-5}$ \\ \hline\hline
    \end{tabular}
\end{table}

The three BPs used in our numerical analysis of the baryon asymmetry $Y_{\Delta B}$ and DM relic density $\Omega_{\rm DM}h^2$ [cf.~Fig.~\ref{fig:scan1}] are collected in Table~\ref{tab:BP2}. These are obtained by implementing our model in {\tt SARAH 4}~\cite{Staub:2013tta} and after checking consistency with all lepton flavor violating constraints using {\tt SPheno 4.0.4}~\cite{Porod:2011nf}.
The observed value of DM relic density is obtained in each case by fixing the Higgs-DM quartic couplings $\lambda_{H \eta}=-\lambda^\prime_{H \eta}$ in Eq.~\eqref{eqn:potential} for a given mass scale $\mu_\eta$ as shown in Table~\ref{tab:BP2}. This assumption is taken in order to ensure the mass of the charged scalar is always higher than the neutral scalar masses (i.e $\eta_R$ and $\eta_I$). All the quartic couplings in Eq.~(\ref{eqn:potential}) not listed in this table are set to be zero.
%The magnitude of the required Higgs-DM quartic coupling increases with the DM mass as the effective freeze-out cross-section goes as $\langle \sigma v \rangle \sim \lambda^2_{H\eta}/m^2_{\eta}$.

We solve the Boltzmann equations~\eqref{eqn:Boltzmann} and \eqref{eqn:Boltzmann2} numerically for the three representative BPs in Table~\ref{tab:BP2}. We assume $
F_{\rm I}  =  F_{\rm II}  =  1/2$
in Eqs.~(\ref{eqn:YN}) and (\ref{eqn:YDelta}), i.e. equal contributions from scotogenic  and type-II seesaw to neutrino masses. This choice  maximizes the $CP$-asymmetry in Eq.~\eqref{eqn:imaginary}, subject to keeping other factors the same. In addition, the ${\cal O}$ matrix is taken to be identity, so that ${\cal O} \widehat{m}_\nu^2 {\cal O}^{\sf T} = \widehat{m}_\nu^2$, and the mass parameter $\mu_{\eta \Delta}$ is assumed to be purely imaginary in Eq.~(\ref{eqn:imaginary}). In doing so, the contribution for the asymmetry coming from the standard decay of $N$'s will not come into play as it requires non-trivial orthogonal matrix $\mathcal{O}$.
%\blue{This is the {\it most conservative} case for the imaginary factor in Eq.~(\ref{eqn:imaginary}), as the factor $F_{\rm I}^{-1} F_{\rm II}^{-1} \geq 4$ and a complex ${\cal O}$ matrix will provide extra CP violation source.}
%The value of $v_\Delta$, new particle masses and quartic couplings for three benchmark points are collected in Table~\ref{tab:BP2}.
The RHNs are taken to be much heavier than the $\eta$ particles to avoid the wash-out of lepton asymmetry from the inverse decay processes $L_\alpha \eta \to N_i$. For the BPs we take, the mass splitting $m_{\eta_{I}} - m_{\eta_{R}}$ (with $\eta_I$ the imaginary part from $\eta^0$) is larger than 100 keV scale, such that the direct detection constraints for inelastic scattering of DM with nucleons~\cite{Chen:2017cqc, Aprile:2017ngb, Suzuki:2018xek} can be evaded.

\begin{figure}[t!]
    \centering
    \includegraphics[height=0.35\textwidth]{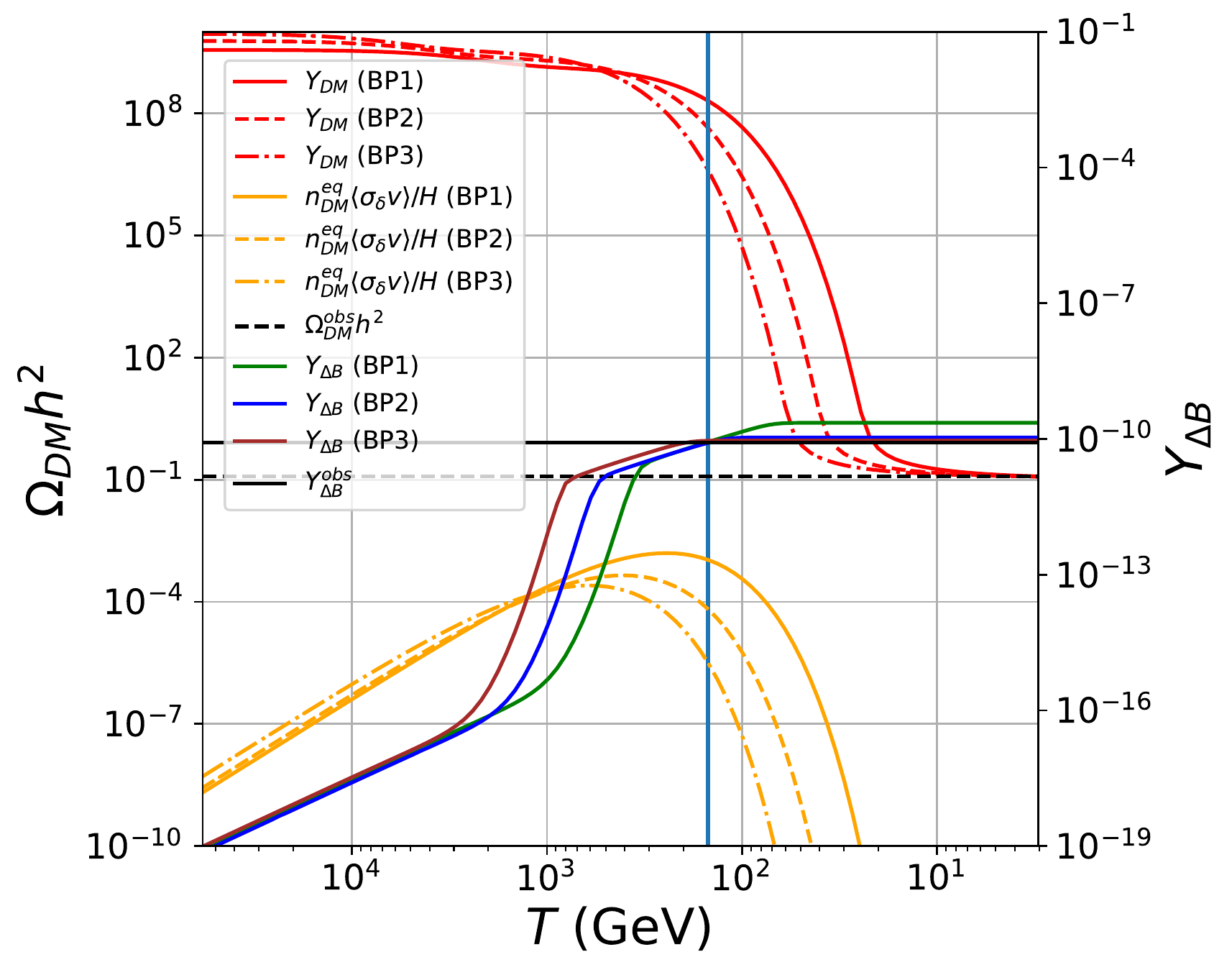}
    \caption{Net baryon number density $Y_{\Delta B}$, DM density $Y_{\rm DM}$ and $n_{\rm DM}^{\rm eq} \langle \sigma v \rangle_\delta/H$ as functions of temperature $T$, for the three BPs in Table~\ref{tab:BP2}.  The solid horizontal black line indicates the observed baryon number density $Y_{\Delta B}^{\rm obs}=(8.718\pm 0.004)\times 10^{-11}$, and the dashed horizontal black line indicates the central value of the observed DM relic density $\Omega_{\rm DM}^{\rm obs}h^2=0.120\pm 0.001$~\cite{Aghanim:2018eyx}. The vertical solid line represents the central value of the sphaleron freeze-out temperature $T_{\rm sph} = (131.7\pm 2.3)$ GeV.
    %Net baryon number density $Y_{\Delta B}$ as function of the $\Delta$-mediator mass, for three different values of $|\mu_{\eta \Delta}|$ (with the argument of $\pi/2$) for each of the three benchmark points in Table~\ref{tab:BP2}.
   }
    \label{fig:lepto}
\end{figure}

The evolutions of the DM relic density $\Omega_{\rm DM}h^2$ and the baryon asymmetry $Y_{\Delta B}$ are evaluated using {\tt micrOMEGAs 5.0}~\cite{Belanger:2018mqt} and the results are presented in Fig.~\ref{fig:lepto}.
%One can clearly see the dependence of baryon asymmetry $\Delta B$ on the mediator masses $m_{N_i}$ and $m_{\Delta^{}}$ as well as the trilinear coupling $\mu_{\eta \Delta}$.
The time evolutions of DM relic density for BP1, BP2, and BP3 are denoted, respectively, by the red solid, dashed and dot-dashed curves.
%The observed value of the relic density is obtained in each case by fixing the Higgs-DM quartic couplings $\lambda^\prime_{\eta H}=-\lambda^{\prime \prime}_{\eta H}=\lambda_{\eta H}$ in Eq.~\eqref{eqn:potential} for a given mass scale ($\mu_\eta$) as shown in Table~\ref{tab:BP2}. The magnitude of the required Higgs-DM quartic coupling increases with the DM mass as the effective freeze-out cross-section goes as $\langle \sigma v \rangle \sim \lambda^2_{H\eta}/m^2_{\eta}$.
For each choice of the DM mass, the maximal contribution to baryon asymmetry comes in the vicinity of the $s$-channel resonance in Fig.~\ref{fig:asymfeyn}, i.e when $2m_\eta \to  m_\Delta$. In Fig.~\ref{fig:lepto} we have fixed the $\Delta$-mediator mass at the resonance point and have satisfied the required baryon asymmetry by fixing the trilinear coupling $\mu_{\eta \Delta}$ as shown in Table~\ref{tab:BP2}.
%We find that the size of \blue{$\mu_{\eta \Delta}$ needed for baryon asymmetry decreases as the DM mass increases. }
%There also exist solutions where the observed baryon asymmetry can be obtained away from the resonance peak, as seen in Fig.~\ref{fig:scan1}.

%The dependence of baryon asymmetry on the trilinear coupling $\mu_{\eta\Delta}$ and the triplet scalar mass $m_\Delta$ is shown in Fig.~\ref{fig:scan1}.
%For each of the three BPs given in Table~\blue{S1}, we show the variation of $Y_{\Delta B}$ as function of the mediator mass for  different values of $\mu_{\eta \Delta}$, as shown by the solid, dashed and dot-dashed lines with green, blue and red, which correspond respectively to BP1, BP2, and BP3. The enhancement of baryon asymmetry at the resonance $m_{\Delta} \simeq 2m_\eta$ can be clearly seen. As expected in Eq.~(\ref{eqn:imaginary}), increasing the amplitude $|\mu_{\eta\Delta}|$ results in a larger baryon asymmetry. %We have fixed $|\mu_{\eta\Delta}|$ for each BP in Table~\ref{tab:BP2} to be the minimum value for which the observed baryon asymmetry can be obtained at the resonance.
%As seen in Fig.~\ref{fig:scan1}, with larger trilinear couplings, one can also achieve the observed asymmetry away from the resonance point.

\begin{figure}[t!]
\centering
\includegraphics[width=0.4\textwidth]{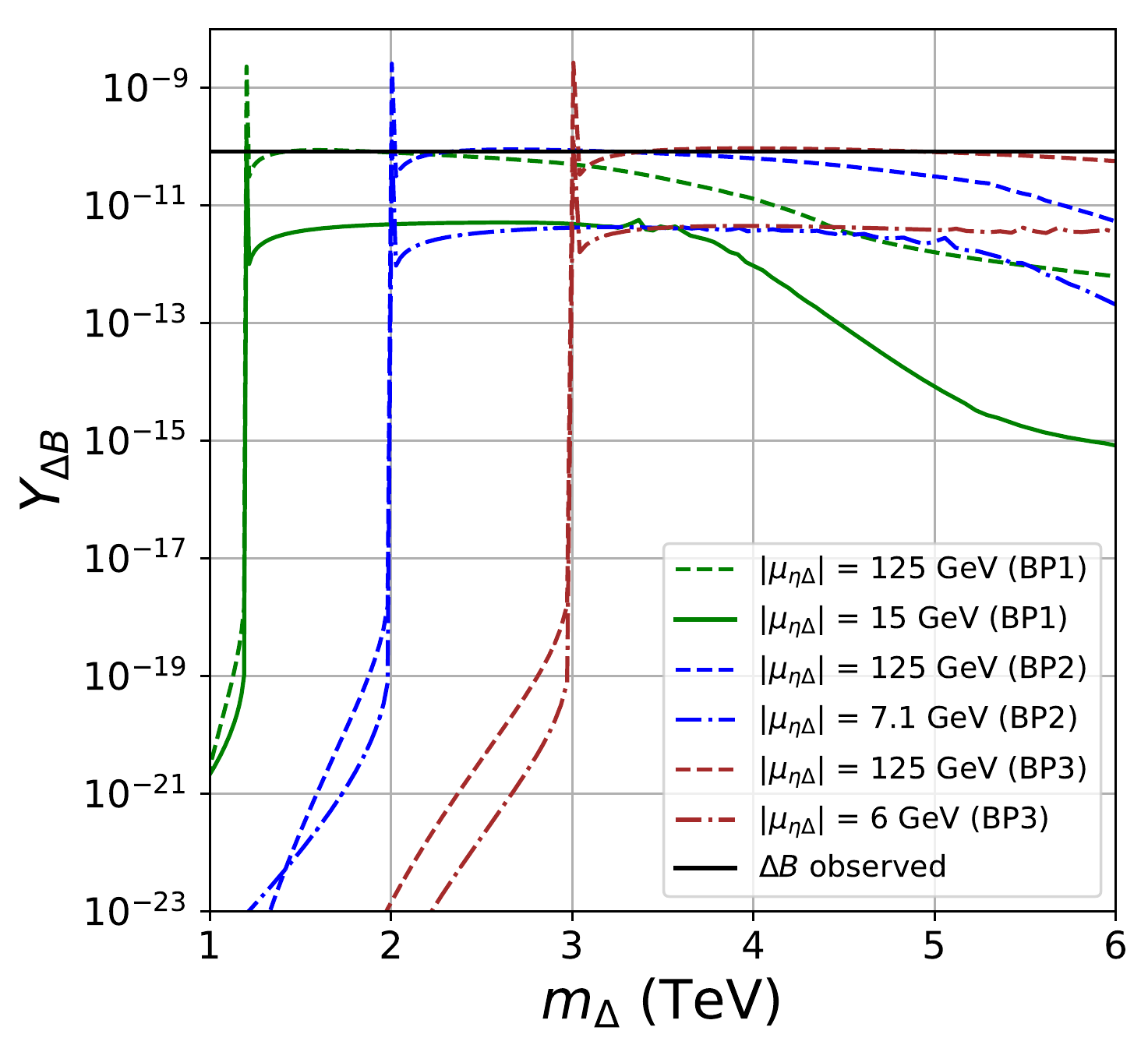}
\caption{Net baryon number density $Y_{\Delta B}$ as function of the $\Delta$-mediator mass, for three different values of $|\mu_{\eta \Delta}|$ (with the argument of $\pi/2$) for each of the three BPs in Table~\ref{tab:BP2}. The solid horizontal black line indicates the central value of the observed baryon number density $Y_{\Delta B}^{\rm obs}=(8.718\pm 0.004)\times 10^{-11}$.
}
\label{fig:scan1}
\end{figure}

The dependence of baryon asymmetry on the absolute value of the trilinear coupling $|\mu_{\eta\Delta}|$ and the triplet scalar mass $m_\Delta$ is shown in Fig.~\ref{fig:scan1}. For each of the three BPs given in Table~\ref{tab:BP2}, we show the variation of $Y_{\Delta B}$ as function of the mediator mass for  different values of $\mu_{\eta \Delta}$, as shown by the solid, dashed and dot-dashed lines with green, blue and red, which correspond respectively to BP1, BP2, and BP3. In the numerical calculations we have included the resonance effect in Eq.~(\ref{eqn:delta2}) when the center-of-mass energy is close to the triplet scalar mass $m_\Delta$.
%[@Arnab, more details?]}
The enhancement of baryon asymmetry at the resonance $m_{\Delta} \simeq 2m_\eta$ can be clearly seen in Fig.~\ref{fig:scan1}. Note that in the narrow-width approximation, the thermal-averaged cross section $\langle \sigma v\rangle_\delta$ asymptotically reaches a finite value [cf.~Eq.~\eqref{eq:asyNWA}] which determines the height of the peak in Fig.~\ref{fig:scan1}, whereas the sharp drop right after the resonance is due to the Boltzmann suppression in Eq.~\eqref{eq:asyNWA}.
As expected in Eq.~(\ref{eqn:imaginary}), increasing the magnitude of  $\mu_{\eta\Delta}$ results in a larger baryon asymmetry. We have fixed $|\mu_{\eta\Delta}|$ for each BP in Table~\ref{tab:BP2} to be the minimum value for which the observed baryon asymmetry can be obtained at the resonance. Note that for larger trilinear couplings, one can also achieve the observed asymmetry away from the resonance point.
%In the right panel of Fig.~\ref{fig:lepto}, %where we vary the mediator mass we see
%there is a slight bump around half of the RHN mass for lower values of $\mu_{\eta \Delta}$ ($\sim 1$ GeV). This is due to mutual cancellation between the \emph{s}- and \emph{t}-channel contributions, which in turn lowers the wash-out rate and as a result enhance slightly the baryon asymmetry. As can be inferred from Fig.~\ref{fig:lepto}, we can get the observed baryon asymmetry of $10^{-10}$~\cite{Aghanim:2018eyx} for suitable choice of the $\mu_{\eta\Delta}$ parameter.
%This is due to the plateau be slight bump around half of the RHN mass for lower values of $\mu_{\eta \Delta}$ ($\sim 1$ GeV).
This plateau region is due to a mutual cancellation between the \emph{s}- and \emph{t}-channel contributions in Fig.~\ref{fig:asymfeyn}, which in turn lowers the wash-out rate, and as a result, slightly enhances  the baryon asymmetry.
%
%(right panel) we have shown the correlation of the mediator mass and the asymmetry in which one may notice that there is a spike in the asymmetry at every resonance point after which it reaches a plateau and then gradually dies down. Now, we notice that the overall feature shifts upward as we increase the trilinear coupling which is trivial as it strengthens the production of the asymmetry as shown in eq.\eqref{eqn:delta2}. But, one may also notice that if we take the trilinear coupling sufficiently high then there is a whole range of mediator mass which one can satisfy the desired leptonic asymmetry.

\begin{figure*}[t!]
    \centering
   % \begin{tabular}{lr}
    \includegraphics[width=0.45\textwidth]{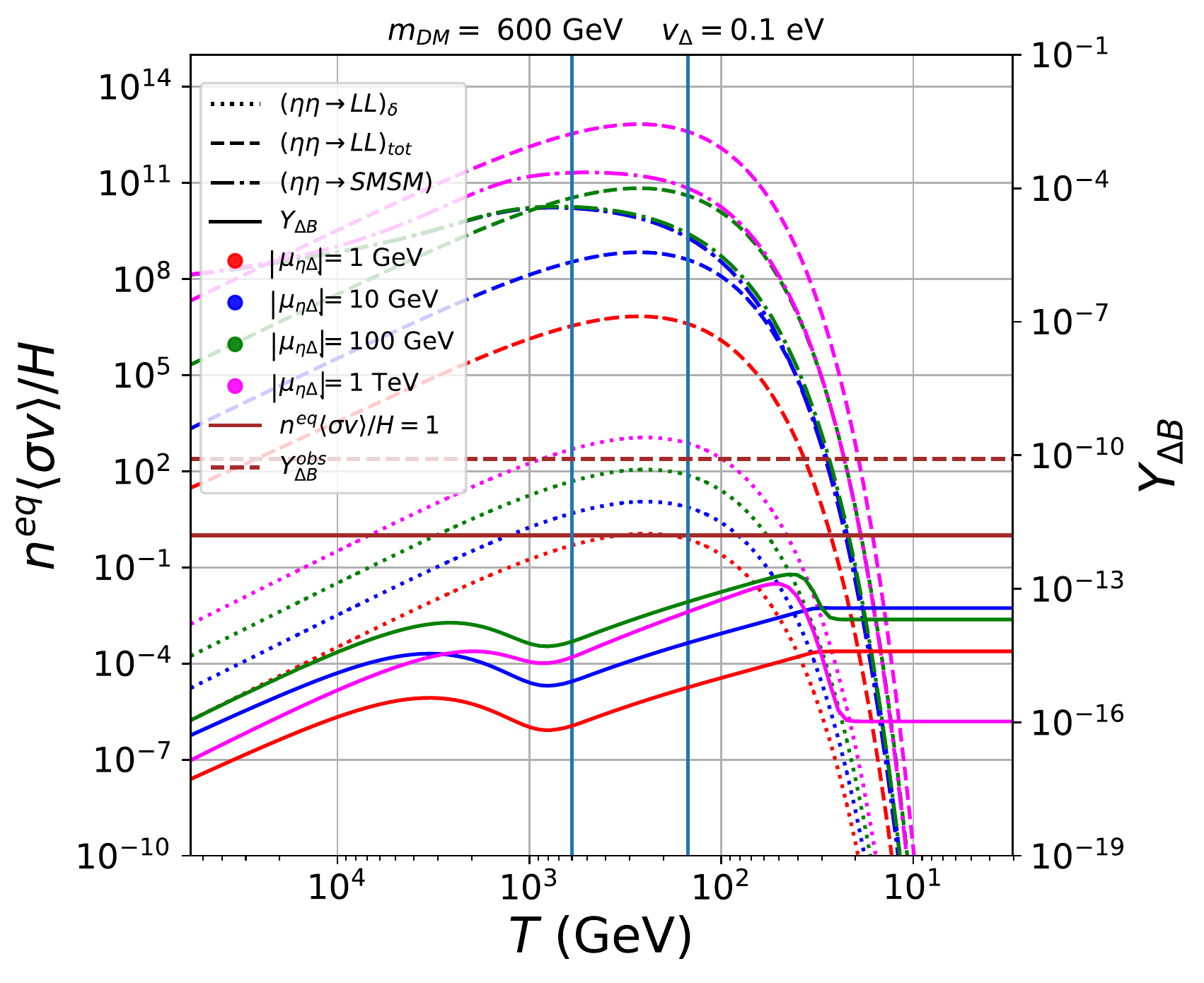}
    \includegraphics[width=0.45\textwidth]{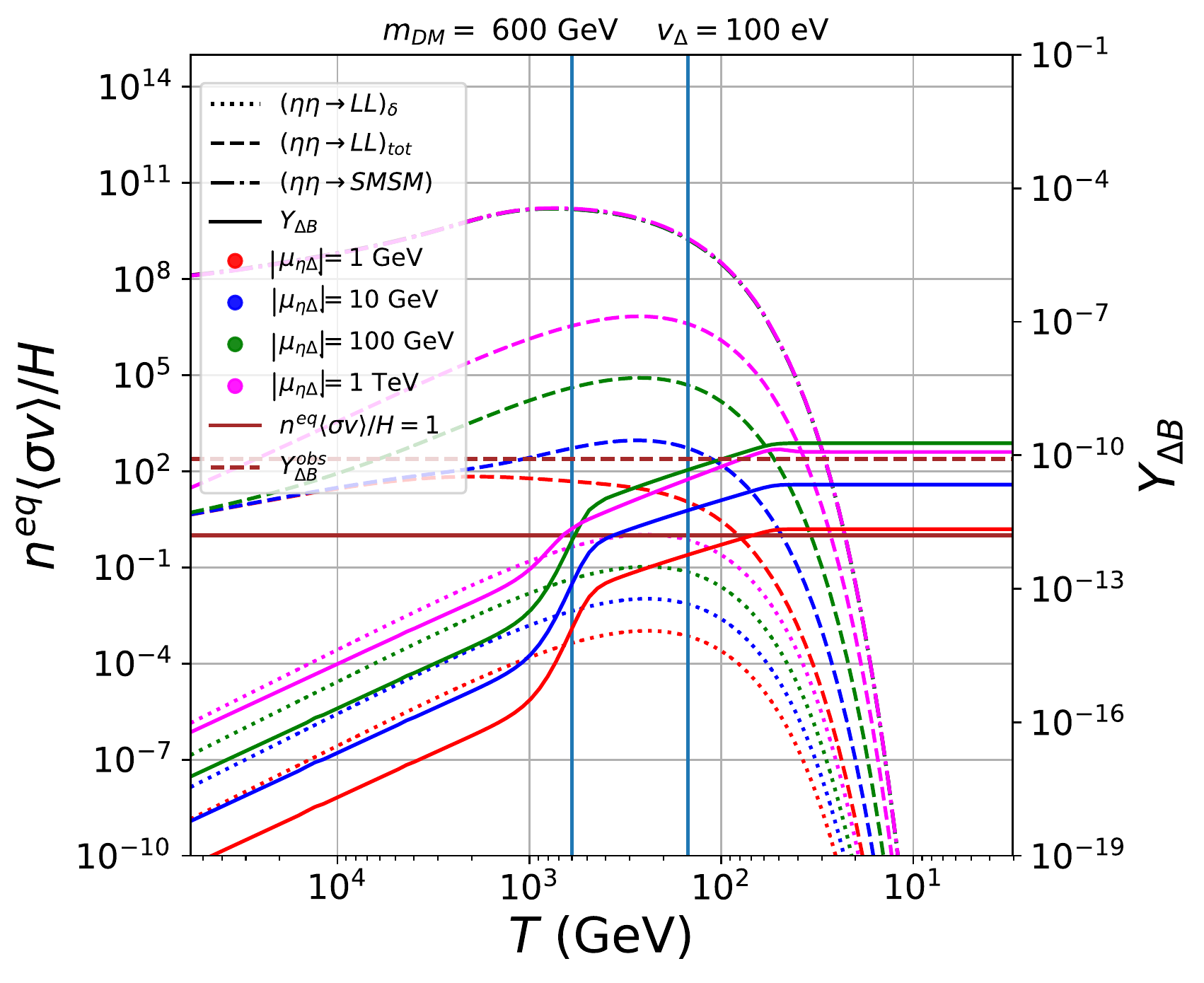}
    %\includegraphics[width=0.5\textwidth]{scan.pdf}
     % \end{tabular}
    \caption{Net baryon number density $Y_{\Delta B}$ and $n^{\rm eq} \langle \sigma v \rangle/H$ for the processes $(\eta\eta \to LL)_\delta$, $(\eta\eta \to LL)_{\rm tot}$ and $(\eta \eta \to {\rm SM} \, {\rm SM})$ as functions of temperature $T$, for the values of $|\mu_{\eta\Delta}| = 1$ GeV, 10 GeV, 100 GeV and 1 TeV for the BP1 in Table~\ref{tab:BP2}, with $v_\Delta = 0.1$ eV (left) and $v_\Delta = 100$ eV (right).  The solid horizontal black line indicates the central value of the observed baryon number density $Y_{\Delta B}^{\rm obs}=(8.718\pm 0.004)\times 10^{-11}$, and the dashed horizontal black line indicates the central value of the observed DM relic density $\Omega_{\rm DM}^{\rm obs}h^2=0.120\pm 0.001$. The vertical solid line represents the central value of the sphaleron freeze-out temperature $T_{\rm sph} = (131.7\pm 2.3)$ GeV.}
    \label{fig:scan2}
\end{figure*}

The dependence of baryon asymmetry on the triplet VEV $v_\Delta$ can be seen in the plots in Fig.~\ref{fig:scan2}. Here the solid lines are for the evolution of $Y_{\Delta B}$ as function of $T$, and the dotted, dashed and dot-dashed lines denote the evolution of the thermally-averaged cross sections $\langle \sigma v \rangle$, respectively for the processes $(\eta\eta \to LL)_\delta$, $(\eta\eta \to LL)_{\rm tot}$ and $(\eta \eta \to {\rm SM} \, {\rm SM})$. The red, blue, green and magenta lines are respectively for the BPs with $|\mu_{\eta \Delta}| = 1$ GeV, 10 GeV, 100 GeV and 1 TeV, and the left and right panels are respectively for the VEVs of $v_\Delta = 0.1$ eV and $v_\Delta = 100$ eV. Other parameter are set to be the same as for BP1. When the VEV $v_\Delta$ gets larger, the Yukawa coupling $Y^\Delta  \propto  v_\Delta^{-1}$ will be smaller, so the wash-out effect will be suppressed and the resultant baryon asymmetry $Y_{\Delta B}$ will be larger, as can be seen by comparing the left and right panels of Fig.~\ref{fig:scan2}.
%However, when $v_\Delta$ is too large, say at the MeV scale or higher, the Yukawa coupling $Y^\Delta$ will be too small to generate the observed baryon asymmetry. The upper and lower bound on $v_\Delta$ from leptogenesis for BP1 is presented in Fig.~\ref{fig:scan}.
%it is clear that the scenarios with $v_\Delta = 10$ eV could lead to larger baryon asymmetry than those with $v_\Delta = 0.1$ eV.

\begin{figure}[!t]
    \centering
    \includegraphics[width=0.45\textwidth]{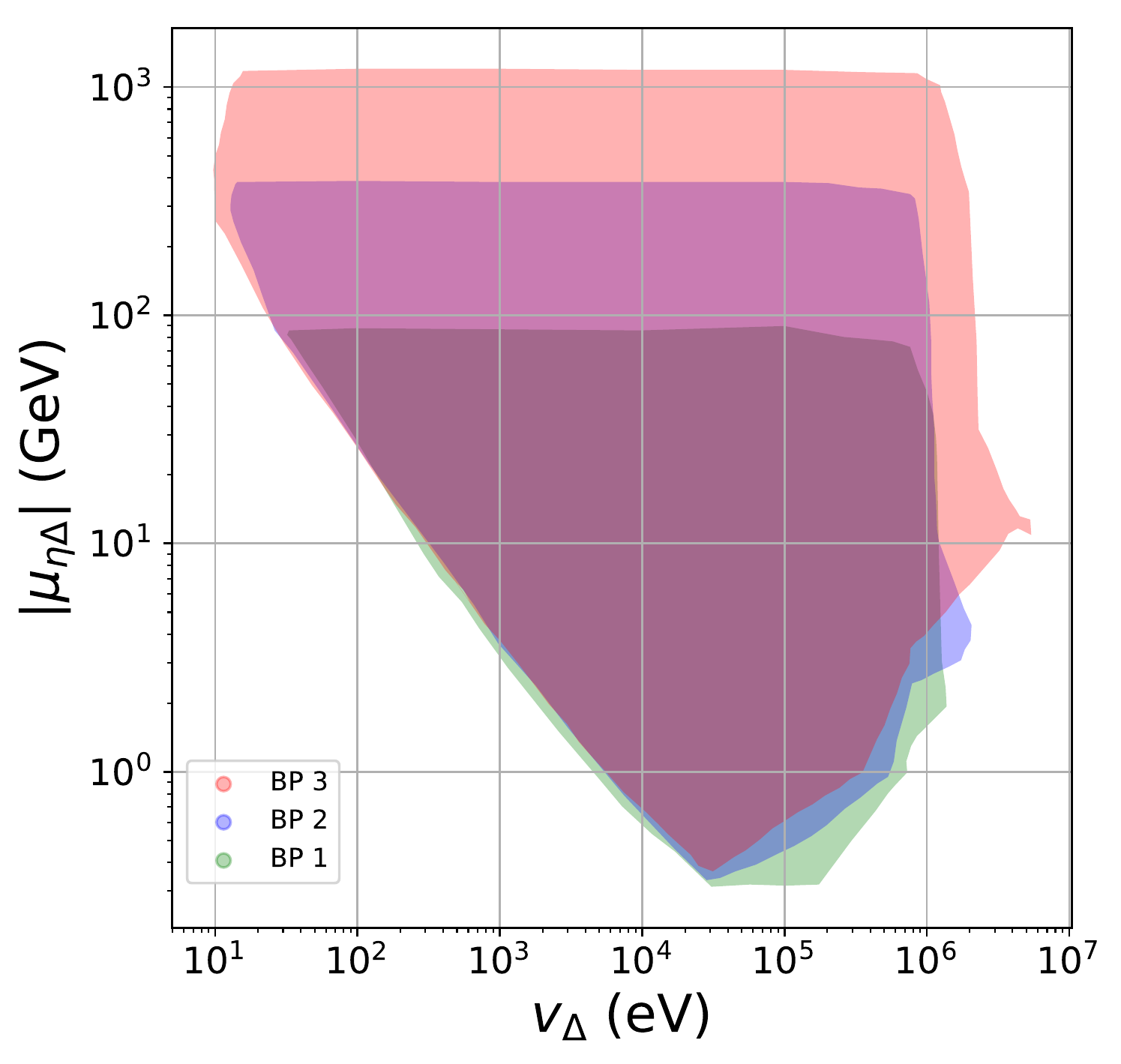}
    \caption{Allowed parameter space of $v_\Delta$ and $|\mu_{\eta \Delta}|$ for the three BPs simultaneously satisfying the observed relic density and baryon asymmetry, with all other relevant parameters set to be the same as in Table~\ref{tab:BP2}.}
    %The red and blue shaded regions \GR{satisfy the observed DM relic abundance and baryon asymmetry, respectively}. Thus, the overlap of these two shaded regions is the allowed parameter space in this model. One would also notice a tail near the end which is due the dependence of the asymmetry $\propto \mu_{\eta \Delta}/v_\Delta$, now as the mass of the DM increases the tails shortens due the suppression from the heavy neutrino.}\BD{(i) Use $Y_{\Delta B}$ instead of $\Omega_B$ (ii) Are you using only the central values or the $1\sigma$ range for the observed values? (iii) Why is there a tail for the blue regions? (iv) What is $\lambda_5$ here?}}
    \label{fig:scan}
\end{figure}

%In the model we are considering, the baron asymmetry crucially depends on the trilinear coupling $\mu_{\eta \Delta}$ and the triplet VEV $v_\Delta$, as exemplified in Figs.~\ref{fig:scan1} and \ref{fig:scan2}.
Varying $v_\Delta$ and $|\mu_{\eta\Delta}|$ and fixing all other relevant parameters as in Table~\ref{tab:BP2},
we obtain the allowed regions of $v_\Delta$ and $|\mu_{\eta\Delta}|$ which simultaneously satisfy the observed baryon number density $Y_{\Delta B}^{\rm obs}=(8.718\pm 0.004)\times 10^{-11}$ and  the observed DM relic density $\Omega_{\rm DM}^{\rm obs}h^2=0.120\pm 0.001$, as shown by the shaded regions in Fig.~\ref{fig:scan} for the three BPs.\footnote{To find the maximum allowed parameter space for $Y_{\Delta B}$, we take the maximal $CP$ phase for $\mu_{\eta\Delta}$ and keep all the points with $Y_{\Delta B}\geq Y_{\Delta B}^{\rm obs}$. Since the $CP$-asymmetry depends on Im($\mu_{\eta\Delta}$), cf. Eq.~\eqref{eqn:delta2}, we can always adjust the $CP$ phase accordingly for a fixed $|\mu_{\eta\Delta}|$ to get $Y_{\Delta B}=Y_{\Delta B}^{\rm obs}$. Therefore, we only show $|\mu_{\eta\Delta}|$ and not its phase in Fig.~\ref{fig:scan}.}
%The shaded regions in Fig.~\ref{fig:scan} satisfy both the observed results.
%It is worth noticing that the dominant process \textcolor{blue}{(i.e $\eta \eta \rightarrow W W$)} responsible for DM relic abundance is independent of $\mu_{\eta\Delta}$  and $v_{\Delta}$, whereas
%the subdominant process \textcolor{blue}{(i.e $\eta \eta \rightarrow L L$)} weakly depends on them, which leads to the upper limits for each $v_{\Delta}$ when we impose the observed $\Omega_{\rm DM}$.
 From this parameter scan, we find both lower and upper bounds on the triplet VEV and the trilinear coupling for each BP:
\begin{align}
{\rm BP1}: & \;\; 40 \, {\rm eV} \lesssim v_\Delta \lesssim 1.5 \, {\rm MeV}, \quad 0.3 \, {\rm GeV} \lesssim |\mu_{\eta \Delta}| \lesssim 80 \, {\rm GeV}, \nonumber \\
{\rm BP2}: & \;\; 20 \, {\rm eV} \lesssim v_\Delta \lesssim 1.2 \, {\rm MeV}, \quad 0.3 \, {\rm GeV} \lesssim |\mu_{\eta \Delta}| \lesssim 380 \, {\rm GeV}, \nonumber \\
{\rm BP3}: & \;\; 10 \, {\rm eV} \lesssim v_\Delta \lesssim 20 \, {\rm MeV}, \quad 0.3 \, {\rm GeV} \lesssim |\mu_{\eta \Delta}| \lesssim 1.2 \, {\rm TeV}.
\end{align}
%$30 \, {\rm eV} \lesssim v_\Delta \lesssim 20$ MeV.
The lower limit on $|\mu_{\eta\Delta}|$ is set by the $Y_{\Delta B}$ requirement, while the upper limit is set by the relic density requirement, which is governed by the $\eta\eta \to WW$ process that is independent of $v_\Delta$ for most part of the parameter space. As for the limits on $v_\Delta$,
when $v_\Delta$ is very small, the Yukawa coupling $Y^\Delta \propto v_\Delta^{-1}$ is so large that the washout effect from $LL\to \eta\eta$ is too strong. On the other hand, when $v_\Delta$ is very large the coupling $Y^\Delta$ is too small to produce sufficient baryon asymmetry. Although one would expect that the suppression of $Y^\Delta$ can be compensated by increasing $\mu_{\eta \Delta}$ [cf.~Fig.~\ref{fig:asymfeyn} (left)], this in turn increases the mass splitting between $\eta_R$ and $\eta_I$, which decreases $Y^N$ [cf.~Fig.~\ref{fig:asymfeyn} (right)] to maintain the neutrino mass. Due to this reason there is a sharp upper bound on $v_\Delta$.

The leptogenesis mechanism in this model can thus be directly tested at future lepton colliders by measuring the Yukawa couplings of doubly-charged scalars~\cite{Dev:2018upe}, which is intimately related to $v_\Delta$ and the active neutrino masses and mixings via Eq.~(\ref{eqn:YDelta}). This model can also be tested at future hadron and lepton colliders by searches of the beyond SM particles in the scotogenic and type-II seesaw models, as detailed in Section~\ref{sec:collider}.
%More details can be found in the {\it Supplemental Material}.

%The effect of the trilinear coupling $\mu_{\eta \Delta}$ and the mediator mass \blue{$m_{\Delta}$} on the baryon asymmetry is further illustrated in the right panel of Fig.~\ref{fig:lepto}.

%\BD{Discuss (i) Lower limit on the DM mass, (iii) Small $\mu$ and vacuum stability. (iv) Resonance feature in right panel.  (vi) Maximal $F$ gives minimum $\delta$. Also maximal phase for trilinear.}

%the lower values of $\mu_{\eta \Delta}$ ($\mu_{\eta \Delta} = 15$ GeV for $m_{\eta}= 600$ GeV and $\mu_{\eta \Delta} = 7$ GeV for $m_{\eta}= 1$ TeV) at the resonance region i.e $M_{\Delta} = 2m_{\eta}$. But, the region opens up for higher values of the trilinear coupling but, it has an upper bound from the requirement of the stability of the potential.

\section{Collider signals}
\label{sec:collider}

The BPs chosen for our model to simultaneously explain baryogenesis, DM and neutrino masses involve TeV-scale beyond SM scalars and heavy RHNs which can be directly tested at current and future high-energy colliders. For instance, the neutral and charged triplet scalars can be directly searched for at the Large Hadron Collider (LHC)~\cite{Aaboud:2017qph, CMS-PAS-HIG-16-036}, as well as in future High-Luminosity LHC (HL-LHC)~\cite{Perez:2008ha, Melfo:2011nx, Mitra:2016wpr, Dev:2018kpa, Antusch:2018svb}, 100 TeV hadron colliders~\cite{Du:2018eaw, Dev:2018kpa} and lepton colliders~\cite{Agrawal:2018pci}, or indirectly probed in the high-precision low-energy experiments like MOLLER~\cite{Dev:2018sel}. It is important to note here that the allowed range of $v_\Delta$ in Fig.~3 correspond to the Yukawa couplings $5\times 10^{-9} \lesssim Y^\Delta \lesssim 3\times 10^{-3}$ which give rise to prompt dilepton signals in the $\Delta^{\pm\pm}$ decays for the triplet masses considered in Table~\ref{tab:BP2}.

The charged $\eta^\pm$ scalars can be produced in association with the neutral DM particle $\eta^0$ through the $W$ boson, i.e. $pp \to W^\ast \to \eta^\pm \eta^0 \to \eta^0 \eta^0 W^{(\ast)}$~\cite{Wan:2018eaz}. The inert doublet scalars can also be produced from their couplings to the SM $Z$ boson via $pp \to \eta_R \eta_I j$ or the SM Higgs through $pp \to \eta^0 \eta^0 j$ (with $j$ being an energetic jet)~\cite{Avila:2019hhv}.
%With the DM $\eta^0$ escaping the detector without leaving any trace, we will have an (off-shell) W boson plus large missing energy at colliders~\cite{Wan:2018eaz}.
The inert doublet sector can then be constrained by the mono-$W$~\cite{Sirunyan:2017jix, Aaboud:2018xdl} and monojet~\cite{Khachatryan:2014rra, Aaboud:2017phn} searches at the LHC. Our model can in principle be distinguished from the pure scotogenic or pure type-II seesaw model at colliders using both inert doublet and triplet scalar signatures.
%from simultaneous observations of signatures of presence of If the heavy RHNs, inert and triplet scalars can all be found at future high-energy colliders, then our model can be distinguished from the pure scotogenic or type-II seesaw models.

In the scotogenic model, the RHNs do not mix directly with the light active neutrinos.
%, or otherwise they would contribute to active neutrino masses at tree-level and to the baryon asymmetry via the standard leptogenesis mechanism.
For our chosen BPs, the heavy RHNs are heavier than the inert doublet and can only be produced at high-energy colliders from the off-shell decay $\eta^{\pm \, \ast} \to \ell_\alpha^\pm N_i$, followed by $N_i \to \ell_\alpha^\pm \eta^{\mp\,(\ast)} \to \ell_\alpha^\pm \eta^0 W^{\mp \, (\ast)}$.  Due to the Majorana nature of the heavy RHNs, we can get same-sign dileptons, as in the Keung-Senjanovi\'{c} process~\cite{Keung:1983uu}, but now with significant missing transverse energy (MET) due to the presence of $\eta^0$ in the final state. The SM background for $\ell^\pm \ell^\pm+W+$MET is expected to be higher than that without MET, and a detailed simulation is needed to estimate the prospects of RHN signals in this model at future colliders.
%~\cite{Keung:1983uu} from heavy neutrino production and decay at high-energy colliders~\cite{Dicus:1991fk, Pilaftsis:1991ug, Datta:1993nm, Han:2006ip, delAguila:2007qnc, Atre:2009rg, Panella:2001wq, Alva:2014gxa, Kang:2015uoc, Degrande:2016aje, Dev:2019rxh}, just as in the standard type-I seesaw~\cite{Minkowski:1977sc, Mohapatra:1979ia, Yanagida:1979as, GellMann:1980vs, Glashow:1979nm} and left-right symmetric models~\cite{Pati:1974yy, Mohapatra:1974gc, Senjanovic:1975rk}. If the heavy RHNs, inert and triplet scalars can all be found at future high-energy colliders, then our model can be distinguished from the pure scotogenic or type-II seesaw models.

%{\bf Conclusion.--}
\section{Conclusion}
\label{sec:conclusion}

We have proposed a new technique to generate the observed baryon asymmetry of the universe only via the tree-level interference of $2\to 2$ scatterings or $1\to3$ decays.
%, without the requirement of loops at the TeV-scale.
%This is possible when we have at least two subprocesses in the $2\to2$ scatterings or $1\to{n}$ (with ${n} \geq 3$) decays.
%This is possible only when we go for the genesis of the asymmetry coming from $2\rightarrow 2$ process or $1\rightarrow 3$ process.
The $CP$-violating asymmetry comes from the absorptive part of the propagators.
%Now, this technique as a possibility of asserting the origin of the asymmetry from the discovery of the mediator particle. Which can be probed in the next-generation colliders.
We have illustrated this mechanism explicitly in a simple scotogenic model with type-II seesaw, in which the asymmetry is generated in the $\Delta L=2$ processes $\eta \eta \to L L$ mediated by $s$-channel triplet scalars and $t$ or $u$-channel RHNs. The neutrino masses receive contributions from both scotogenic and  type-II seesaw mechanisms. The real part of the neutral component of the inert doublet $\eta$ serves as a DM candidate. As shown in Fig.~\ref{fig:lepto} the baryon asymmetry and DM relic density are correlated and both can be matched to their observed values for (sub-)TeV inert doublet and triplet masses. The baryogenesis requirements impose both lower and upper bounds on the triplet VEV, as shown in Fig.~\ref{fig:scan}, with testable consequences at future colliders.

%In this model the Standard Model is extended by a scalar doublet and three copies of RHNs both \emph{odd} under $Z_2$ giving mass to neutrino radiatively. Additionally there is a scalar triplet which also gives mass to neutrino but at tree level.
%We have shown that the asymmetry can be achieved with the imaginary part of the trilinear coupling where the absolute value can be as low as $\mu_{\eta \Delta}\sim 15$ GeV for Dark matter to be around $m_\eta \sim 600$ GeV.
%The Yukawa's taken to satisfy the neutrino mass and leptonic asymmetry are well within the present LFV (Lepton Flavor Violation) bounds.

\acknowledgments

%{\bf Acknowledgments.--}
We thank Pei-Hong Gu for comments on the draft. AD thanks the Particle Theory Group at Washington University in St. Louis for warm hospitality, where this work was initiated. YZ thanks Kaladi Babu for useful discussions, and the High Energy Theory Group at Oklahoma State University for warm hospitality, where this work was completed. The work of PSBD and YZ was supported in part by the U.S. Department of Energy under Grant No.  DE-SC0017987 and in part by the MCSS. This  work  was  also supported in part by the Neutrino Theory Network Program under Grant No. DE-AC02-07CH11359. The work of SKK and AD was supported by the NRF of Korea Grants No. 2017K1A3A7A09016430, and No. 2017R1A2B4006338.

\appendix

\section{Scalar potential and scalar masses}
\label{sec:scalars}

The most general scalar potential for the SM Higgs doublet $H \equiv (H^+,\, H^0)$, inert doublet $\eta \equiv (\eta^+,\, \eta^0)$ and triplet $\Delta \equiv (\Delta^{++},\, \Delta^+,\, \Delta^0)$ is given by
%\begingroup\makeatletter\def\f@size{9}\check@mathfonts
\begin{align}
\label{eqn:potential}
    V  = &
    - \mu^2_H (H^\dagger H) + \mu^2_\eta (\eta^\dagger \eta)
    - \mu^2_\Delta {\rm Tr} \left[\Delta^\dagger \Delta\right] \nonumber \\
   & + (\mu_{H\Delta} \widetilde{H}^\dagger \Delta H
    + \mu_{\eta \Delta} \eta^\dagger \Delta^\dagger \widetilde{\eta} + {\rm H.c.}) \nonumber \\
   & + \lambda_H (H^\dagger H)^2
     + \lambda_\eta (\eta^\dagger \eta)^2
    + \lambda_\Delta  \{ {\rm Tr}\left[\Delta^\dagger \Delta\right] \}^2  \nonumber \\
    &+ \lambda^\prime_\Delta {\rm Tr} \left[\Delta^\dagger \Delta \Delta^\dagger \Delta \right]  + \lambda_{H\eta} |H^\dagger \eta|^2 + \lambda^\prime_{H\eta} (H^\dagger H)(\eta^\dagger \eta) \nonumber \\
    & + \lambda^{\prime \prime}_{H\eta} ((H^\dagger \eta)^2 + {\rm H.c.})
    + \lambda_{H\Delta} (H^\dagger H) {\rm Tr} \left[\Delta^\dagger \Delta\right] \nonumber \\
    &+ \lambda^\prime_{H\Delta} {\rm Tr} [H^\dagger \Delta \Delta^\dagger H]
    + \lambda_{\eta\Delta} (\eta^\dagger \eta) {\rm Tr} \left[\Delta^\dagger \Delta\right] \nonumber \\
     &+ \lambda^\prime_{\eta \Delta} {\rm Tr}[\eta^\dagger \Delta \Delta^\dagger \eta] \,,
\end{align}
%\endgroup
where $\widetilde{H} = i\sigma_2 H^\ast$, $\widetilde{\eta} = i\sigma_2 \eta^\ast$ and the mass parameters $\mu_{H,\, \eta,\, \Delta}^2>0$ so that both $H$ and $\Delta$ obtain non-vanishing VEVs, i.e. $\langle H^0 \rangle = v \simeq 246$ GeV and $\langle \Delta^0 \rangle = v_\Delta$. To generate the baryon asymmetry from the tree-level interference effect, the coupling $\mu_{\eta\Delta}$ is assumed to be complex, and all other parameters in Eq.~(\ref{eqn:potential}) are assumed to be real.

The physical masses for the neutral and charged scalars can be obtained from minimization of the scalar potential in Eq.~(\ref{eqn:potential}).
%which is detailed in the Appendix.
Note that here the doublet $\eta$ is odd under the $Z_2$ symmetry, which is essential to provide a DM candidate, and does not mix the SM Higgs and the triplet.
%which is necessary for the neutral real component $\eta_{\rm R}$ to be a stable DM candidate.
%All the scalar masses in the scotogenic plus type-II seesaw model with the SM Higgs $H$, the inert doublet $\eta$ and the triplet $\Delta$ can be obtained from the scalar potential (\ref{eqn:potential}).
In particular, when we take the first-order derivative of the potential with respect to the VEVs $v$ and $v_\Delta$,  the  solutions of the tadpole equations for \{$\mu^2_H,\mu^2_\Delta$\} are given by
\begin{align}
    \mu^2_H & \ = \ \frac{1}{2}(\lambda_H v^2 - 2\sqrt{2}\mu_{H\Delta}v_\Delta + \lambda_{H\Delta}v^2_\Delta) \,, \\
    \mu^2_\Delta & \ = \ \frac{1}{2}(\lambda_\Delta v^2_\Delta - 2\sqrt{2} \frac{\mu_{H\Delta}}{v_\Delta} v^2 + \lambda_{H\Delta}v^2 ) \,.
\end{align}
After replacing \{$\mu^2_H,\mu^2_\Delta$\} in the scalar potential, the mass matrix for the real scalars reads
\begin{align}
    {\cal M}^0 & \ = \ \begin{pmatrix} \lambda_H v^2 & -\sqrt{2}\mu_{H\Delta}v \\
   -\sqrt{2}\mu_{H\Delta}v & \frac{\mu_{H\Delta}v^2}{\sqrt{2}v_\Delta} \end{pmatrix} \,,
\end{align}
from which we can get two mass eigenvalues for the real component from $H^0$ and the scalar $\Delta^0_R$ which is the real part of $\Delta^0$. In the case of $\mu_{H \Delta} \sim \mathcal{O}(100)$ keV, the two CP-even scalar masses turn out to be
\begin{eqnarray}
m_h^2 \simeq \lambda_H v^2 \,, \quad m_{\Delta^0_R}^2 \simeq \frac{\mu_{H\Delta}v^2}{\sqrt{2}v_\Delta} \,,
\end{eqnarray}
with the first one ($h$) identified as the SM-like Higgs boson. The masses of the pseudo-scalar and the charged scalars from the triplet are respectively
\begin{align}
m_{\Delta_I}^2 & \ = \ \frac{\mu_{H\Delta}}{\sqrt{2}v_\Delta}(v^2 + 4v^2_\Delta) \,, \\
m_{\Delta^\pm}^2 & \ = \ m_{\Delta^{\pm\pm}}^2 = \left(\frac{\mu_{H\Delta}}{\sqrt{2}v_\Delta}+ \frac{1}{4}\lambda^\prime_{H\Delta}\right)(v^2 + 2v^2_\Delta) \,.
\end{align}
Finally the masses for real scalar $\eta_{R}$, pseudo-scalar $\eta_{I}$ and the charged scalars $\eta^\pm$ from the $Z_2$-odd doublet $\eta$ are respectively
\begin{align}
    m_{\eta_{R,I}}^2 \ = \ & \frac{1}{2}\left[2\mu^2_\eta + (\lambda_{H\eta} + \lambda^\prime_{H\eta} \pm \lambda^{\prime \prime}_{H\eta})v^2 \right. \nonumber \\
    &+ \left. (\lambda_{\eta \Delta}v_\Delta \mp 2\sqrt{2}|\mu_{\eta \Delta}|) v_\Delta \right] \,, \\
    m_{\eta^\pm}^2 \ = \ & \frac{1}{2}\left(2\mu^2_\eta + \lambda_{H\eta}v^2 \right.
    + \left. (\lambda_{\eta \Delta} + \lambda^\prime_{\eta \Delta}) v^2_\Delta \right) \,.
\end{align}

\section{\bf Thermal cross sections}
\label{sec:xsec}

The general expression for the thermally-averaged cross section for the processes in Eqs.~(\ref{eqn:Boltzmann}) and (\ref{eqn:Boltzmann2}) is~\cite{Belanger:2018mqt}
%\begin{widetext}
\begin{align}
\label{eq:sigv}
& \langle \sigma v \rangle(i_1 i_2 \rightarrow f_1 f_2) \ = \
\frac{1}{2T m^2_{i_1} K_2(m_{i_1}/T)m^2_{i_2} K_2(m_{i_2}/T)} \nonumber \\
& \times \int^\infty_{s_{\rm in}}
\int^1_{-1}\frac{1}{32\pi}\frac{|\mathcal{M}|^2}{\sqrt{s}}p_{i_1 i_2}p_{f_1 f_2}K_1(\sqrt{s}/T) \ {\rm d} s\  {\rm d} \cos\theta \,,
\end{align}
%\end{widetext}
where $T$ is the temperature, $K_{i}$ the modified Bessel functions of order $i$, $\mathcal{M}$ is the amplitude for the process $i_1 i_2 \rightarrow f_1 f_2$, and
\begin{eqnarray}
p_{ij} & \ \equiv \ & \frac12 \sqrt{{\lambda(s,m^2_i,m^2_j)}/{s}} \,, \label{eq:B2} \\
\quad s_{\rm in} & \ \equiv \ & {\rm max} [(m_{i_1} + m_{i_2})^2,(m_{f_1} + m_{f_2})^2] \,, \label{eq:B3} \\
\lambda(x,y,z) & \ \equiv \ & x^2 + y^2 + z^2 + 2xy + 2xz + 2yz \, .
\end{eqnarray}
In Eq.~(\ref{eqn:Boltzmann2}), $\langle \sigma v\rangle_{\rm tot}(\eta \eta \rightarrow L L)$ and $\langle \sigma v\rangle^{}(\eta \overline{L} \rightarrow \eta L)$ are respectively for the amplitudes:
\begin{widetext}
\begin{align}
\left| \mathcal{M}_{\rm tot}(\eta \eta \rightarrow L L) \right|^2 \ = \ &  \frac{\widehat{m}^2_\nu}{v^2_\Delta}\frac{F^{2}_{\rm I} |\mu_{\eta \Delta}|^2s}{(s - m^2_\Delta)^2 + m^2_\Delta\Gamma^2_\Delta}
+ \sum_i F^{2}_{\rm II} \frac{\widehat{m}^2_\nu}{\Lambda^2_{ii}}m^2_{N_i}s\left[\frac{1}{t - m^2_{N_i}} + \frac{1}{u - m^2_{N_i}}\right]^2 \nonumber \\
&+ \frac{F_{\rm I}^{} F_{\rm II}^{} |\mu_{\eta \Delta}|(s - m^2_\Delta)}{(s - m^2_\Delta)^2 + m^2_\Delta\Gamma^2_\Delta}
\sum_i \frac{\widehat{m}^2_\nu}{\Lambda_{ii} v_\Delta} m_{N_i}s\left[\frac{1}{t - m^2_{N_i}} + \frac{1}{u - m^2_{N_i}}\right] \,, \\
\left| \mathcal{M}(\eta \overline{L} \rightarrow \eta L) \right|^2 \ = \ & \frac{\widehat{m}^2_\nu}{v^2_\Delta}\frac{F^{2}_{\rm I} |\mu_{\eta \Delta}|^2(m^2_\eta - t)}{(t - m^2_\Delta)^2}
+ \sum_i \frac{\widehat{m}^2_\nu}{\Lambda^2_{ii}}\frac{F^{2}_{\rm II}m^2_{N_i}s}{(s - m^2_{N_i})^2 + m^2_{N_i}\Gamma^2_{N_i}} \nonumber \\
&+ \frac{F_{\rm I}^{} F_{\rm II}^{} |\mu_{\eta \Delta}| (m^2_\eta - t)}{(t - m^2_\Delta)^2}\sum_i \frac{\widehat{m}^2_\nu}{\Lambda_{ii} v_\Delta} m_{N_i}s\left[\frac{s - m^2_{N_i}}{(s - m^2_{N_i})^2 + m^2_{N_i}\Gamma^2_{N_i}}\right] \,.
\end{align}
\end{widetext}
The cross section $\langle \sigma v \rangle (\eta \eta \to {\rm SM}\, {\rm SM})$ in Eq.~(\ref{eqn:Boltzmann}) can be found in~\cite{Barbieri:2006dq,LopezHonorez:2006gr}, with ``SM SM'' referring to the all the possible channels involving the quarks, leptons, scalar and gauge bosons in the SM.

\section{\bf Maximum Asymmetry}
\label{sec:appasym}

In Eq.~(\ref{eqn:Boltzmann2}) $\langle \sigma v \rangle_\delta(\eta \eta \rightarrow L L)$ is for the amplitude given in Eq.~(\ref{eqn:delta2}) which produces the lepton asymmetry. In the narrow-width approximation, the asymmetry in Eq.~(\ref{eqn:delta2}) at the resonance point simplifies to
\begin{align}
\label{eq:asymapp}
    \delta & \ \simeq \ 4
    \sum_{i} \: {\rm Im}
\left[\mu_{\eta \Delta} \big\{Y^N Y^{\Delta^*}(Y^{N})^{\sf T} \big\}_{ii} \right] \nonumber \\
&\times s m_{N_i}  \delta(s-m^2_\Delta)\left[\frac{1}{t-m^2_{N_i}} + \frac{1}{u-m^2_{N_i}}\right] \, ,
\end{align}
where we have used
\begin{align}
    \frac{m_\Delta \Gamma_{\Delta}}{(s-m^2_\Delta)^2 + m^2_\Delta \Gamma^2_\Delta} \ \xrightarrow[]{\Gamma_\Delta/m_\Delta\to 0} \ \pi \delta(s-m^2_\Delta).
\end{align}
Plugging the amplitude~\eqref{eq:asymapp} back into Eq.~\eqref{eq:sigv} and integrating over $s$, we get
\begin{widetext}
\begin{align}
    \langle \sigma v\rangle_\delta(\eta \eta \rightarrow LL) & \ = \  \frac{z}{128\pi^2m_\eta^5K^2_2(z)}\int^\infty_{s_{\rm in}}ds\frac{p_{\eta \eta}p_{L L}}{\sqrt{s}}K_1\left(\frac{\sqrt{s}z}{m_\eta}\right)\delta \nonumber \\
    & \simeq \ \frac{4\pi}{m^2_\eta}\frac{r^4_\Delta}{\widetilde{\mu}_{\eta \Delta}^2\sum_{\alpha,\beta}|Y_{\alpha \beta}|^2}\Gamma_{\Delta\rightarrow \eta \eta}\Gamma_{\Delta\rightarrow LL}
\sum_{i} \: {\rm Im}
\left[\mu_{\eta \Delta} \big\{Y^N Y^{\Delta^*}(Y^{N})^{\sf T} \big\}_{ii} \right]\frac{K_1(r_\Delta z)}{r_{N_i}K^2_2(z)} \, , \label{eq:asyNWA}
\end{align}
\end{widetext}
where $\widetilde{\mu}_{\eta \Delta} = \mu_{\eta \Delta}/m_\Delta$, $r_i = m_i/m_\eta$, $p_{ij}$ and $s_{\rm in}$ are defined in Eqs.~\eqref{eq:B2} and \eqref{eq:B3} respectively, and
\begin{align}
    \Gamma_{\Delta \to \eta\eta} \ & = \ \frac{1}{16\pi} \widetilde{\mu}_{\eta\Delta}^2  p_{\eta\eta} \, , \\
    \Gamma_{\Delta \to LL} \ & = \ \frac{1}{16\pi}|Y_{\alpha\beta}|^2 p_{LL} \, .
\end{align}
It can be seen from Eq.~\eqref{eq:asyNWA} that the thermally-averaged asymmetry asymptotically
reaches a finite value at the resonance, which determines the height of the peak in Fig.~\ref{fig:scan1}.

\bibliographystyle{utphys}
\bibliography{ref}

\end{document}